\documentclass[preprint]{aastex61}
\usepackage{verbatim}
\usepackage{graphicx}
\usepackage[space]{grffile}
\usepackage{latexsym}
\usepackage{textcomp}
\usepackage{multirow,booktabs}
\usepackage{amsfonts,amsmath,amssymb}
\usepackage{natbib}
\usepackage{url}
\usepackage{hyperref}
\usepackage{verbatim}
\usepackage{longtable}
\usepackage{float}
\usepackage[caption = false]{subfig}
\usepackage{enumitem}
\usepackage{listings}
\usepackage{media9}
\setlist[enumerate]{itemsep=-2mm}

\newif\iflatexml\latexmlfalse
\DeclareGraphicsExtensions{.pdf,.PDF,.png,.PNG,.jpg,.JPG,.jpeg,.JPEG}

\usepackage[utf8]{inputenc}
\usepackage[ngerman,english]{babel}

\usepackage{babel}

\shortauthors{Catherine Zucker}

\begin{document}


\title{Mapping Distances Across the Perseus Molecular Cloud Using CO Observations, Stellar Photometry, and Gaia DR2 Parallax Measurements}


\author{Catherine Zucker}
\affil{Harvard Astronomy, Harvard-Smithsonian Center for Astrophysics, 60 Garden St., Cambridge, MA 02138, USA}

\author{Edward F. Schlafly}
\affil{Lawrence Berkeley National Laboratory, One Cyclotron Road, Berkeley, CA 94720, USA}

\author{Joshua S. Speagle}
\affil{Harvard Astronomy, Harvard-Smithsonian Center for Astrophysics, 60 Garden St., Cambridge, MA 02138, USA}

\author{Gregory M. Green}
\affil{Kavli Institute for Particle Astrophysics and Cosmology, Physics and Astrophysics Building, 452 Lomita Mall, Stanford, CA 94305, USA}

\author{Stephen K. N. Portillo}
\affil{Harvard Astronomy, Harvard-Smithsonian Center for Astrophysics, 60 Garden St., Cambridge, MA 02138, USA}

\author{Douglas P. Finkbeiner}
\affil{Harvard Astronomy, Harvard-Smithsonian Center for Astrophysics, 60 Garden St., Cambridge, MA 02138, USA}

\author{Alyssa A. Goodman}
\affil{Harvard Astronomy, Harvard-Smithsonian Center for Astrophysics, 60 Garden St., Cambridge, MA 02138, USA}

\defcitealias{Green_2018}{G18}



\begin{abstract}
We present a new technique to determine distances to major star-forming regions across the Perseus Molecular Cloud, using a combination of stellar photometry, astrometric data, and $\rm ^{12} CO$ spectral-line maps. Incorporating the Gaia DR2 parallax measurements when available, we start by inferring the distance and reddening to stars from their Pan-STARRS1 and 2MASS photometry, based on a technique presented in \citet{Green_2014, Green_2015} and implemented in their 3D ``Bayestar" dust map of three-quarters of the sky. We then refine the Green et al. technique by using the velocity slices of a CO spectral cube as dust templates and modeling the cumulative distribution of dust along the line of sight towards these stars as a linear combination of the emission in the slices. Using a nested sampling algorithm, we fit these per-star distance-reddening measurements to find the distances to the CO velocity slices towards each star-forming region. This results in distance estimates explicitly tied to the velocity structure of the molecular gas. We determine distances to the B5, IC348, B1, NGC1333, L1448, and L1451 star-forming regions and find that individual clouds are located between $\approx 275-300$ pc, with typical combined uncertainties of $\approx 5\%$. We find that the velocity gradient across Perseus corresponds to a distance gradient of about 25 pc, with the eastern portion of the cloud farther away than the western portion. We determine an average distance to the complex of $294\pm 17$ pc, about 60 pc higher than the distance derived to the western portion of the cloud using parallax measurements of water masers associated with young stellar objects. The method we present is not limited to the Perseus Complex, but may be applied anywhere on the sky with adequate CO data in the pursuit of more accurate 3D maps of molecular clouds in the solar neighborhood and beyond. 
\end{abstract}

\section{Introduction}
As the most active site of star formation within the solar neighborhood \citep[$\lesssim$ 300 pc,][]{Bally_2008}, the Perseus Molecular Cloud complex has been the subject of a wealth of continuum and spectral-line observations in recent years. These studies have targeted both the global properties of the molecular cloud and the properties of discrete, high-density pockets of gas and dust in which small groups or clusters of stars are forming (e.g. L1451, L1448, NGC1333, B1, IC348, and B5; see Figure \ref{fig:perseus_regions}). Distance estimates to these well-known star-forming regions show a wide degree of dispersion---varying anywhere between 200-350 pc---and are based on a variety of techniques. Nevertheless, accurate distance measurements are critically important for constraining properties like clump mass or star formation efficiency of the gas across Perseus, whose proximity to the sun facilitates high-resolution observations of the star formation process.  

 \citet{Cernis_1990, Cernis_1993} use interstellar extinction to find distances to both the eastern (IC348 at 260 pc) and western portions (NGC1333 at 220 pc) of Perseus. Because optical light is extinguished by molecular clouds, \citet{Cernis_1990, Cernis_1993} determine photometric distances to unextinguished foreground and extinguished background stars, thereby constraining the distance of the jump in extinction. More recently, \citet{Hirota_2008, Hirota_2011} determine distances to the western portion of Perseus, by obtaining trigonometric parallax measurements of water masers associated with young stellar objects embedded in the NGC1333 and L1448 regions---finding a distance of 235 pc ($\varOmega=4.25\pm0.32 \; \rm mas$, where $\varOmega$ is hereafter parallax) for the former and 232 pc ($\varOmega=4.31\pm0.33 \; \rm mas$) for the latter. Adopting a slightly different technique, \citet{Lombardi_2010} calculates a distance to Perseus by comparing the density of low-extinction foreground stars \citep[determined by the NICEST color excess method;][]{Lombardi_2009} with the prediction for foreground stellar density from the Besan\c{c}on Galactic model \citep{Robin_2003}, finding a distance of 260 pc to B1 and B5 and 212 pc to L1448. 
 
\citet{Schlafly_2014} is the first to systematically map distances across the entire Perseus Complex. \citet{Schlafly_2014} first obtains distances and reddenings to batches of stars in over a dozen sightlines throughout Perseus using optical photometry \citep[as outlined in][]{Green_2014}. Then, \citet{Schlafly_2014} adopts a model whereby the stellar distances and reddenings are caused by a single dust screen, to which they find the distance using a Markov Chain Monte Carlo (MCMC) analysis. The \citet{Schlafly_2014} results suggest a distance gradient, but often with large uncertainties on individual lines of sight ($\approx 10-20\%$) and with farther distances, overall, than suggested in \citet{Hirota_2008,Hirota_2011}, typically lying around 260-315 pc. 
 
One possible explanation for the distance discrepancies is that the Perseus Complex consists of clouds at several distances along the line of sight, a scenario bolstered by the large velocity gradient ($\rm \approx 5-10 \; km \; s^{-1}$) observed across the cloud in CO \citep{Ridge_2006}. However, unlike for clouds outside the solar neighborhood, this velocity gradient cannot be mapped to a distance gradient via a Galactic rotation curve \citep[by the so-called ``kinematic distance" method, see][]{Roman_Duval_2009} because the distance resolution of the kinematic method is coarser than the typical peculiar motions for objects so close to the Sun. 

In this analysis, we build upon the work of \citet{Schlafly_2014} and present a new technique to map velocities to distances across the Perseus Molecular Cloud, by combining information on the spatial distribution of CO emission with distance and reddening estimates towards thousands of stars obtained from \citet{Green_2018} (hereafter \citetalias{Green_2018}). By using the velocity slices of a CO spectral cube as dust templates and modeling the cumulative distribution of dust along the line of sight as a linear combination of optical depth in CO velocity slices, we perform a Monte Carlo analysis to determine which distance configuration for the velocity slices is most consistent with the distance and reddening estimates to sets of stars in regions across Perseus.

In \S \ref{data}, we introduce the key data sets in this analysis, including the photometry used to derive the stellar distance and reddening estimates, the astrometric data used to inform the stellar distances, and the CO spectral-line data used as a dust tracer. In \S \ref{green_method}, we briefly explain how the stellar distance and reddening estimates are derived from the photometric catalogs discussed in \S \ref{data}. In \S \ref{targets}, we discuss our target selection and the batch of stars we consider towards each region of interest. In \S \ref{model}, we present our Bayesian model for the line-of-sight distribution of dust as a function of the CO velocity slices. In \S \ref{MCMC}, we discuss the nested sampling algorithm we use to perform the parameter estimation. In \S \ref{results}, we present our distance estimates to the CO slices for major star-forming regions across the Perseus Molecular Cloud and compare our results with distance estimates from the literature. We discuss the implications of our results in \S \ref{discussion} and conclude in \S \ref{conclusion}. 

\section{Data} \label{data}
Determining distances to molecular clouds based on the distribution of their CO emission is a two-step process. In the first step, we obtain the \textit{per-star} posterior probability density function (PDF) of distance and reddening for stars towards the Perseus Molecular Cloud, based on their optical (Pan-STARRS1, hereafter PS1) and near-infrared (2MASS) photometry. When available, we fold in existing knowledge on the distance to each star, based on its Gaia DR2 parallax measurement. In the second step, we use the slices of a CO spectral cube as dust templates---by multiplying each slice by a gas-to-dust conversion coefficient---and sample for the probable range of distances to each velocity slice given the distance and reddening estimates to stars towards the same region. In this section, we briefly describe the key data sets necessary for this analysis. 

\subsection{Pan-STARRS1 Photometry} \label{ps1}
The Pan-STARRS1 $3\pi$ survey is a deep optical survey of the three quarters of the sky north of $\delta=-30^\circ$ \citep{Chambers_2016}. The PS1 observations are obtained using a 1.8-meter telescope situated on Mount Haleakala in Hawaii, which has been equipped with a Gigapixel camera with a $3^\circ$ field of view and pixel scale of 0.258$\arcsec$. The survey observes in five broadband filters, the $grizy_{P1}$ bands, spanning $400-1000$ nm \citep{Chambers_2016}. The images are processed by the PS1 Image Processing Pipeline, which automatically performs photometric and astrometric measurements on the reduced data \citep{Magnier_2016}. The stellar posteriors on distances and reddening we utilize in this work (from \citetalias{Green_2018}; see \S \ref{green_method}) are based on catalog coadds of single-epoch photometry derived from the PS1 DR1 $3\pi$ steradian survey, which reaches typical single-exposure depths of 22.0, 21.8, 21.5, 20.9, and 19.7 magnitudes (AB) in the $g$, $r$, $i$, $z$, and $y_{P1}$ bands, respectively \citep{Chambers_2016}. 

\subsection{2MASS Photometry} \label{2mass}
The Two Micron All Sky Survey (2MASS) is a near infrared survey of the full sky targeting the $J$, $H$, and $K_s$ bandpasses at $1-2 \; \micron$ \citep{Skrutskie_2006}. The 2MASS observations are obtained via two 1.3-meter telescopes located on Mount Hopkins, Arizona and Cerro Tololo, Chile, both of which are equipped with a three-array survey camera with an $8.5\arcmin$ field of view and a pixel scale of 2$\arcsec$ \citep{Skrutskie_2006}. Like PS1, the image processing pipeline automatically performs photometric and astrometric measurements on the reduced images. The resulting catalogs achieve typical $10\sigma$ point-source depths of 15.8, 15.1, and 14.3 magnitudes (Vega) for the $J$, $H$, and $K_s$ bands, with bright sources possessing photometric uncertainties on the order of $< 0.03$ mag. To derive the stellar posteriors on distance and reddening based on \citetalias{Green_2018}, we specifically use the 2MASS ``high-reliability" catalog (see \S 3.3 in \citetalias{Green_2018} for more details). 

\subsection{Gaia DR2 Astrometric Data} \label{gaia}
Gaia DR2 is an intermediate data release from the Gaia mission \citep{Gaia_2016}, and includes proper motion and parallax measurements for one billion stars, accompanied by all-sky broadband optical photometry (in the $G$, $G_{RP}$ and $G_{BP}$ bands) and radial velocity measurements for stars on the bright end ($G < 13$ mag). We only utilize the astrometric catalog \citep{Lindegren_2018} and defer potential incorporation of Gaia photometry to future work. The astrometric catalog has a limiting magnitude of $G=21$ mag and a bright limit of $G=3$ mag, with uncertainties on the brightest end of $\approx 0.04$ mas and on the faintest end of $0.7$ mas. For a full treatment of the astrometric processing pipeline, see \citet{Lindegren_2018}. We implement the same quality cuts for the Gaia data as given in Equation 11 of \citet{Lindegren_2018}. 

\subsection{CO COMPLETE Data} \label{complete}
We employ maps of the $\rm ^{12}CO$ (1-0) transition in Perseus, taken from the COMPLETE Survey of Star Forming Regions\footnote{All the COMPLETE data are publicly available and can be downloaded from the Harvard Dataverse \href{https://dataverse.harvard.edu/dataverse.xhtml?alias=complete}{here}. The corresponding DOI for the $\rm ^{12}CO$ COMPLETE cube of Perseus is doi:10904/10072} \citep{Ridge_2006}. We have chosen the COMPLETE data due to its high-angular resolution (half-power beamwidth $\approx 46"$) and its relatively low noise (mean rms per channel $\approx$ 0.35 K). Its spectral resolution in the $\rm ^{12}CO$ (1-0) line is $\rm 0.06 \; km \; s^{-1}$. The total areal coverage of the COMPLETE survey towards Perseus is 6.25 x 3 deg$^2$. 

We have chosen the $\rm ^{12} CO$ (1-0) line primarily because it is more abundant and has more extended emission, which allows us to target larger areas of the sky (and therefore more stars), enabling better distance estimates. In theory, we could employ maps targeting the rarer CO isotopologues (e.g. $\rm ^{13}CO, C^{18} O$) in lieu of or in addition to the $\rm ^{12}CO$ data. There are several arguments to be made for using the $\rm ^{13}CO$ (1-0) line in particular, most notably that it is optically thinner and tends to be more linear with extinction than the $\rm ^{12} CO$ (1-0) line \citep{Pineda_2008}. According to \citet{Pineda_2008}, the $\rm ^{12} CO$ line typically saturates in Perseus (and thus non-linear with reddening) around an $A(V) \approx 4$ mag, while the $\rm ^{13} CO$ line typically saturates around an $A(V) \approx 5$ mag. Thus, in theory, $\rm ^{13}CO$ allows us to probe one magnitude deeper, and thereby spatially closer to the densest star-forming cores. However, gaining an extra magnitude does not translate to an appreciable increase in the number of stars available for analysis, since few stars are visible behind such a high dust column density.  Moreover, because of the higher critical density of $\rm ^{13} CO$ \citep[it becomes self-shielded at $A(V) \approx 2$ mag, vs. $A(V) \approx 1$ mag for $\rm ^{12} CO$; see Tables 4 \& 6 in][]{Pineda_2008}, it is a comparatively poor dust template in the more diffuse, extended envelopes of the clouds, from which we draw the bulk of our stars.\footnote{For the clouds we target in \S \ref{targets} (see Figure \ref{fig:perseus_regions}), we find that the inclusion of regions with total line-of-sight visual extinction $4 < A(V) < 5$ mag typically only increases our star count by approximately a few dozen to a few hundred, enlarging our sample size by only about 5-15\%. As discussed in \S \ref{targets}, we delineate cloud boundaries using $\rm ^{12} CO$ integrated intensity contours. However, since $\rm ^{13}CO$ is less extended (both spatially and kinematically) we would need to adopt a comparatively higher CO integrated intensity threshold to produce reliable dust templates. In cases like L1448, the adoption of even a generous low-level $\rm ^{13}CO$ integrated intensity contour (roughly coincident with the $A(V) \approx 2$ mag self-shielding threshold for $\rm ^{13} CO$) reduces our star count by a factor of three, because we target less area towards the outer envelopes of the clouds (better traced by $\rm ^{12} CO$), from which most of our stars are drawn.} Thus, all the results presented in this work utilize the $\rm ^{12} CO$ line, towards sightlines where $\rm ^{12}$CO does not saturate.  

\section{Obtaining Stellar Distance and Reddening Estimates} \label{green_method}
Based solely on the PS1 (\S \ref{ps1}) and 2MASS (\S \ref{2mass}) photometry, we obtain stellar distance and reddening estimates for stars across Perseus based on the work of \citetalias{Green_2018}. We then post-process the distance-reddening stellar posteriors to incorporate the Gaia DR2 parallax information. The methodology used to derive the distance and reddening posteriors is given in \citet{Green_2014, Green_2015}. In brief, \citetalias{Green_2018} infers a distance modulus $\mu$, reddening $E$, and stellar type $\Theta$ for a star, where $\Theta$ is parameterized by the star's absolute PS1 $r$-band magnitude, $M_r$, and its metallicity, $\left[ \mathrm{Fe} / \mathrm{H} \right]$. By adopting a set of stellar templates that map the star's intrinsic stellar type to its absolute magnitude in different photometric bands, \citetalias{Green_2018} obtains the following relation for the modeled apparent magnitudes for each star:

\begin{equation}
\vec{m}_{mod}=\vec{M}(M_r,\left[ \mathrm{Fe} / \mathrm{H} \right])+\vec{A}(E,R(V)) + \mu
\end{equation}

\noindent where the vector notation indicates that $\vec{m}_{mod}$, $\vec{M}$, and $\vec{A}$ are considered over a range of passbands (e.g. $\vec{m}_{mod}=[m_{mod, \, g}, m_{mod, \, r}, m_{mod, \,  i} ....$]). Then, fixing the extinction curve and assuming $R(V)=3.3$ \citep{Schlafly_2016}\footnote{The \citet{Schlafly_2016} work does not directly measure $R(V)=A(V)/{E(B-V})$ because their observations are insensitive to the gray component of the extinction vector. Rather, \citet{Schlafly_2016} builds a proxy for $R(V)$ using the quantity $(A_g - A_{W2})/(A_g-A_r)$, where $g$ and $r$ are the Pan-STARRS1 $g$ and $r$ band magnitudes and $W2$ is the WISE band two magnitude. Thus, the $R(V)=3.3$ value we quote is actually the proxy $R(V)$ \citet{Schlafly_2016} calculates for their mean extinction vector. See \S5.3 in \citet{Schlafly_2016} for more details.}, the likelihood of observing a star with apparent PS1 and 2MASS magnitudes $\vec{m}$, assuming independent Gaussian photometric uncertainties $\vec{\sigma}$, is a multivariate normal with mean $\vec{m}_{mod}$ and standard deviation $\vec{\sigma}$, evaluated at $\vec{m}$. Prior knowledge of the number density and metallicity of stars across the Galactic disk and halo \citep{Juric_2008, Ivezic_2008}, as well as on the stellar luminosity function \citep{Bressan_2012}, is also incorporated into the model.\footnote{We refer readers to \citet{Green_2014} for a full treatment of these priors. See their \S 4.2.1 for a description of the number density prior, \S 4.2.2 for the metallicity prior, and  \S4.2.3 for the stellar luminosity prior.} Combining the likelihood function and priors, \citetalias{Green_2018} draws from the posterior distribution function of $\mu$, $E$, and $\Theta$ for individual stars using an affine-invariant ensemble Markov Chain Monte Carlo (MCMC) sampler \citep{Goodman_2010}. While we adopt the same likelihood function and priors as \citetalias{Green_2018}, we infer the stellar posteriors using brute-force grid evaluation rather than MCMC. This has the added benefit of producing the same results every time, and is free from convergence issues inherent in MCMC sampling. 

After marginalizing over the $\Theta$ parameters and taking a kernel density estimate of the samples, we produce a two-dimensional gridded stellar posterior describing the probable range of the distance and reddening to each star over the domain $0 < E(B-V) < 7$ mag and $4 < \mu < 19$ mag. Before running our line-of-sight fits, we post-process the gridded distance-reddening posteriors in order to fold in knowledge on the distance to each star based on its Gaia DR2 parallax measurement, when available. This term is multiplicative along the $\mu$ axis and functionally acts as an additional Gaussian likelihood term of the following form:

\begin{equation}
 \label{eq:posterior}
P(\varOmega | \mu ) =\frac{1}{\sigma_{\varOmega} \sqrt{2 \pi}} \exp{[\frac{ -({\varOmega - 10^{\frac{\mu+5}{-5}}})^2   }{2 \sigma_{\varOmega} ^{2}}]}
\end{equation}

\noindent where $\varOmega$ is the observed Gaia parallax measurement in arcseconds, $\sigma_{\varOmega}$ is the uncertainty on the parallax measurement (also in arcseconds), and $\mu$ is the distance modulus bin in the gridded distance-reddening posterior. Gaia stellar parallaxes are available for 65\% of our sample, with typical fractional parallax errors of $\approx 20\%$ for stars with $\mu < 8$ mag. Folding in these Gaia parallax data has two major effects on our stellar posteriors. First, for stars in the solar neighborhood, it substantially narrows the posteriors along the $\mu$ axis when the uncertainties are low, constraining the distance to the star to a few hundredths of a magnitude in distance modulus. Second, for stars which are farther away but with posteriors that are multi-modal in distance modulus, Gaia can usually discriminate between the two modes, and suppresses the incorrect mode significantly. These Gaia-informed posteriors are the ones we implement in our model (see \S \ref{model}). 

\subsection{A note on $R(V)$} \label{rv}
We adopt a fixed value of $R(V)=3.3$ across the entire Perseus complex. However, there is evidence for grain growth across the Perseus Molecular Cloud \citep{Foster_2013}. \citet{Foster_2013} use a Hierarchical Bayesian approach to examine the extinction curve in both the eastern and western portions of Perseus, and they find a strong correlation between $A(V)$ and $R(V)$, which they interpret as evidence of grain growth occurring once moderate optical depths are reached.  For the range of $A(V)$ we are probing ($A(V) \approx 2-4$ mag), our adopted value of $R(V)$ is generally consistent with the finding of \citet{Foster_2013}. They find that for $A(V)\approx 3$ mag, the typical $R(V)$ is $\approx 3.5$. This is also consistent with initial results from the APOGEE reddening survey, which targets bright red giant stars through dense sight lines along Perseus in order to study the shape of the extinction curve in nearby molecular clouds (Schlafly et al. 2018, in prep.) -- they find that for $A(V) \approx 3$ mag, the mean $R(V)$ is Perseus is consistent with 3.3. 

The adoption of a fixed extinction curve will lead to small systematic uncertainties in our distance determinations. In other dust clouds of similar column density and at similar distances (e.g. Hercules), changing $R(V)$ by 1-2 changes distance modulus $\mu \lesssim 0.1$ mag (Zucker et al. 2018, in preparation). Since most of our power to constrain distance comes from the Gaia parallaxes, this effect should be small, because these measurements are insensitive to $R(V)$. $R(V)$ will additionally have an effect on the inferred reddening, but ultimately we only need enough accuracy to constrain the step in reddening associated with the cloud, which is quite large.  That said, there will be some degeneracy between $R(V)$ and the gas-to-dust coefficients we infer.  To help accommodate small errors in the extinction curve, we have added an uncertainty of $0.05$ mag in E($B-V$) to our reddening estimates, by smoothing our surfaces by this amount along the reddening axis. The distance axis is not smoothed. 

\section{Cloud Selection} \label{targets}
\begin{figure}[h!]
\begin{center}
\includegraphics[width=1.0\columnwidth]{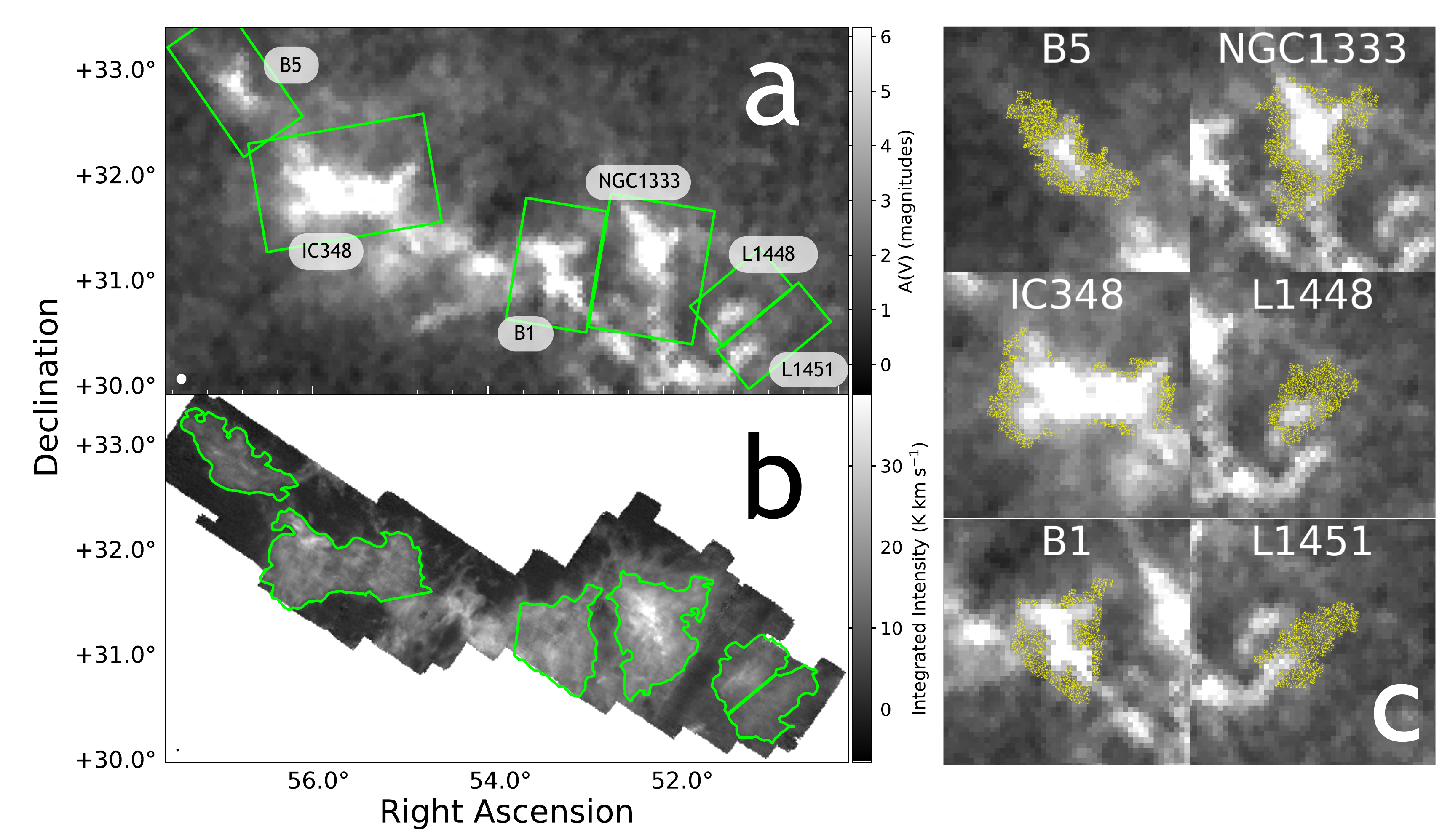}
\caption{{\label{fig:perseus_regions}\textbf{Panel a:} The Perseus Molecular Cloud, as seen in visual extinction \citep{Pineda_2008}. We frame each region of interest (B5, IC348, B1, NGC1333, L1448, and L1451) via the green rectangles. The beam size ($5\arcmin$) is shown in white in the bottom left corner  \textbf{Panel b:} The Perseus Molecular Cloud, as seen in integrated $\rm ^{12}CO$ emission \citep{Ridge_2006}. The beam size ($46\arcsec$) is shown in black in the bottom left corner. To delineate boundaries around the cloud, we apply integrated intensity contours to this map (at $\rm \approx 5-15 \; K \; km \; s^{-1}$, depending on the region) and find the intersection between these contours and the classical regions of extinction, as shown in panel (a). The final boundaries are shown via the green polygons in panel (b). \textbf{Panel c:} The positions of the stars (yellow scatterpoints) considered in our analysis for each region, after masking out lines of sight with high extinction ($A(V) > 4$ mag). An additional cut is also made to remove stars with the worst chi-squared values.}}
\end{center}
\end{figure}

We target every major star-forming region inside the boundaries of the CO COMPLETE survey (see \S \ref{complete}) in this analysis. This includes B5, IC348, B1, NGC1333, L1448, and L1451. We show an extinction map of the Perseus complex in Figure \ref{fig:perseus_regions}a \citep{Pineda_2008} and box these regions with green rectangles, which are apparent as pockets of very high optical depth. We also show a $\rm ^{12} CO$ integrated intensity map of the Perseus complex in Figure \ref{fig:perseus_regions}b \citep{Ridge_2006}. To define boundaries around each cloud we apply integrated intensity contours to the cloud's corresponding $\rm ^{12}CO$ emission ($\rm W(CO) \approx 5-15 \; K \; km \; s^{-1}$, depending on the region).\footnote{The integrated intensity contours are applied to the same moment zero map, computed by integrating over all channels in the $\rm ^{12}CO$ cube provided by the the COMPLETE survey (\texttt{doi:10904/10072}). In reality, only the velocity channels between $\rm -2 \; and \; 12 \; km \; s^{-1}$ have significant emission} In cases where we cannot obtain a closed contour, we find the intersection between a reasonable semi-closed integrated intensity contour and the ``classical" regions of extinction as established in previous studies of Perseus \citep[green rectangles in Figure \ref{fig:perseus_regions}a; see, for instance][]{Rosolowsky_2008, Bally_2008, Kirk_2007,Sadavoy_2014}. These boundaries are \textit{ad hoc}, but they are inclusive of previous definitions of these regions, and we find that our results are robust to modest changes in boundary definition. The final boundaries we use for each region are shown as green polygons in Figure \ref{fig:perseus_regions}b. The Right Ascension and Declination values corresponding to the geometric centroid of each polygon are given in Table \ref{velocity_table}. 

While all the stars we consider lie inside the green polygonal boundaries shown in Figure \ref{fig:perseus_regions}b, we make two additional cut for stars lying inside these regions, based on their total line-of-sight extinction and their chi-squared values. First, at higher reddening, we know that the $\rm ^{12}CO$ line becomes saturated, such that the total integrated CO emission we observe flattens above some extinction threshold \citep[see Figure 6 in][]{Pineda_2008}. Above this threshold, CO can no longer be used as a reliable tracer of dust. 

\citet{Pineda_2008} quantifies the extinction threshold above which $\rm ^{12} CO$ becomes saturated for Perseus. In more detail, \citet{Pineda_2008} combines the $\rm ^{12} CO$ COMPLETE data with a dust extinction map, produced by applying the NICER \citep[Near-Infrared Color Excess method Revised;][]{Lombardi_2001} algorithm to 2MASS photometry, to estimate the total line-of-sight extinction towards Perseus. This NICER extinction map from \citet{Pineda_2008} is the same shown in Figure \ref{fig:perseus_regions}a. \citet{Pineda_2008} finds that on average, $\rm ^{12} CO$ emission in Perseus becomes saturated around $A(V) \approx 4$ mag. Consequently, we mask out all regions inside the green polygons (Figure \ref{fig:perseus_regions}b) with $A(V) > 4$ mag. Specifically, we regrid the NICER extinction map to the same pixel scale as the $\rm ^{12} CO$ COMPLETE cube and exclude all stars that fall inside CO pixels with corresponding NICER $A(V) > 4$ mag.\footnote{The regridding is done via bilinear interpolation} The relative area we mask inside the green polygonal boundaries in Figure \ref{fig:perseus_regions}b is quite high, typically on the order of 25-75\%, so we are, in reality, only probing the outer, lower density envelopes of each region. We mark the positions of stars along unsaturated sightlines inside our boundaries via the gold scatter points in Figure \ref{fig:perseus_regions}c. Finally, we make one additional cut on these stars based on their best-fit chi-squared values, determined via the brute force grid evaluation method described in \S \ref{green_method}. Specifically, we remove the $10\%$ of stars towards each cloud with the lowest chi-squared values, indicating that they are unlikely to be described by our stellar models. For each region, between $\approx$ 2,000 and 3,000 stars are used in the parameter estimation described in \S \ref{MCMC}. 

\begin{deluxetable}{ccccccc}
\colnumbers
\tablecaption{\label{velocity_table}}
\tablewidth{700pt}
\tablehead{\colhead{Cloud Name} & \colhead{RA} & \colhead{Dec} & \colhead{Velocity Range} & \colhead{Downsampled Slice Count} & \colhead{Slice Velocities} & \colhead{Slice Width}\\ \colhead{} & \colhead{$^\circ$} &  \colhead{$^\circ$} & \colhead{$\rm km \; s^{-1}$} & \colhead{} & \colhead{$\rm km \; s^{-1}$} &  \colhead{$\rm km \; s^{-1}$} }
\startdata
L1451 & 51.0 & 30.5 & -1.1 -- 7.4 & 4 & [0.3, 2.3, 4.2, 5.8] & 2.1 \\
L1448 & 51.2 & 30.9 & -2.2 -- 7.9 & 5 & [-0.7, 0.8, 2.9, 4.8, 6.6] & 2.0 \\
NGC1333 & 52.2 & 31.2 & -1.3 -- 10.1 & 6 & [-0.1, 1.7, 3.5, 5.5, 7.3, 8.8] & 1.9 \\
B1 & 53.4 & 31.1 & -0.2 -- 9.6 & 5 & [1.1, 2.8, 4.8, 6.7, 8.2] & 2.0 \\
IC348 & 55.8 & 31.8 & 4.2 -- 11.8 & 4 & [5.4, 7.4, 9.0, 10.4] & 1.9 \\
B5 & 56.9 & 32.9 & 6.7 -- 12.4 & 3 & [8.3, 9.7, 11.0] & 1.9 \\
\enddata
\tablecomments{The velocity of the downsampled CO slices that we model as dust screens in our analysis. In column \textbf{(1)} we list the name of the star-forming region. In column \textbf{(2)} we list the Right Ascension for the cloud, computed using the geometric centroid of each green polygonal boundary shown in Figure \ref{fig:perseus_regions}b. In column \textbf{(3)} we list the Declination for the cloud, computed using the same geometric centroid as in (2). In column \textbf{(4)} we show the velocity range towards each region in which CO emission is not saturated and also above the noise level. Given the native spectral resolution of the CO data \citep[$\rm \approx 0.06 \; km \; s^{-1}$;][]{Ridge_2006} there are typically dozens of velocity slices in this velocity range towards each cloud. As a result, we have downsampled the cube along the spectral axis, and in column \textbf{(5)} we list the number of downsampled velocity slices towards each region. In column \textbf{(6)} we summarize the CO intensity-weighted mean velocity of each downsampled CO slice, computed using only pixels along sightlines where the CO does not saturate (see \S \ref{targets} and \S \ref{ds}). In column \textbf{(7)} we indicate the spectral width of the downsampled slices towards each region; to maintain uniformity across the complex, the spectral width is chosen to be close to $\rm 2 \; km \; s^{-1}$ while covering the desired velocity range for each cloud with an integer number of slices.}
\end{deluxetable}

\section{Model} \label{model}
Our model for the cumulative reddening in E($B-V$) along the line of sight out to distance modulus $\mu$---hereafter called the ``reddening profile"---is denoted by: 

\begin{equation}
{\rm E}_{B-V}(\mu; \vec{\alpha})
\end{equation}

where $\vec{\alpha}$ is some set of parameters describing the reddening profile. As in \citetalias{Green_2018}, the posterior probability density of our $\vec{\alpha}$ parameters is determined by the product of the integral over $\mu$ through the individual stellar posterior density functions (derived from the set of PS1 and 2MASS photometry \{$\vec{m}$\}) following ${\rm E}_{B-V}(\mu; \vec{\alpha})$: 

\begin{equation} \label{pdf_function}
p(\vec{\alpha} \mid \{\vec{m}\}) \propto p(\vec{\alpha}) \; \prod_i^{\rm nstars} \int p(\mu_i, E_{B-V}(\mu_i; \vec{\alpha}) \mid \vec{m_i}) \; d\mu_i
\end{equation}

\noindent where $p(\vec{\alpha}$) constitutes our priors and the product of the integral through the individual stellar posterior arrays is our likelihood function. This is simply Bayes' rule, modulo the normalizing constant in the denominator and assuming independence among stars. 

\citetalias{Green_2018} parameterizes the reddening profile as a piecewise linear function in distance modulus, so $\vec{\alpha}$ is given by a set of parameters denoting the increase in reddening, $\Delta {\rm E}_{B-V}$, in fixed distance bins equally spaced in distance modulus between $\mu=4$ mag (63 pc) and $\mu=19$ mag (63 kpc). Here, however, we do not sample the increase in reddening directly or in fixed distance bins along the entire line of sight. Rather, for an individual star lying in CO pixel $j$ we parameterize Perseus' contribution to the line-of-sight reddening profile towards this pixel as the sum of the CO emission in a set of $n$ velocity slices $\{v_{1},v_{2}...v_{n}$\} that lie at distances $\{d_1,d_2...d_n$\} and whose corresponding CO emission intensities $\{I_{j,1},I_{j,2}...I_{j,n}$\} can be converted to reddening by multiplying by a set of gas-to-dust coefficients $\{c_1,c_2,...c_n$\}. The distances are the primary free parameters of interest, to be constrained by the model fit. Since the CO emission along the line of sight differs for stars in different pixels, the reddening profile will vary from star to star. 

We also assume that there is an angularly uniform foreground dust cloud, whose reddening is not accounted for in the total CO emission given by the velocity slices. We parameterize the foreground dust cloud as lying at a distance $d_{fore}$, and with a reddening contribution in $\Delta E_{B-V}$ given by $E_{fore}$. For a star $i$ in CO pixel $j$, the reddening profile is parameterized as follows:

\begin{equation}
{\rm E}_{B-V}(\mu_i; \vec{\alpha})={\rm E}_{B-V}(\mu_i; d_{fore}, E_{fore}, \{d_1,d_2...d_n\}, \{I_{j,1},I_{j,2},...I_{j,n}\}, \{c_1,c_2,...c_n\}),
\end{equation}

where our free parameters are $d_{fore}$, $E_{fore}$, $\{d_1,d_2...d_n\}$, and $\{c_1,c_2,...c_n\}$. The profile takes the form of a step function. Assuming that our velocity slices are ordered by distance ($d_1 < d_2 ... < d_n$), the total line-of-sight reddening out to a distance $d$ for a single star, coincident with CO pixel $j$, would be: 

$$
E_{B-V}= \left\{
        \begin{array}{lr}
            0 & d \leq d_{fore} \\
            E_{fore}& d_{fore} \leq d \leq d_1 \\
            E_{fore} + I_{j,1} \; c_1  & d_1 \leq d \leq d_2 \\  
            ... & \\
            E_{fore} + \sum_{k=1}^{n} \; I_{j,k} \; c_k & d > d_{n}
        \end{array}
    \right.
$$

The assumption of the velocity slices increasing monotonically in distance (such that  $v_1 < v_2 ... < v_n$, so $d_1 < d_2 ... < d_n$) is only for notational purposes above, and is not actually imposed when we perform our line-of-sight fits. In actuality, the velocity slices have the freedom to switch their order. 

In Figure \ref{fig:example_profile}, we show a cartoon line-of-sight reddening profile for a very simple one-slice model (lying at velocity $v_1$), where the CO emission towards our region of interest is described by a single distance component $d_1$. In such a model, we can parameterize the total line-of-sight reddening profile as having a jump in reddening ${E_{fore}}$ at distance $d_{fore}$, plus a second jump in reddening at distance $d_1$, the magnitude of which is given by the CO emission in slice $v_1$ coincident with the star, multiplied by some gas-to-dust coefficient $c_1$. The free parameters defining our reddening profile are $d_{fore}$, $E_{fore}$, $d_1$, and $c_1$, which are fixed across all stars. The only variation in the reddening profile from star to star stems from the magnitude of the reddening jump at distance $d_1$, which is dependent on the CO pixel $j$ coincident with each star on the plane of the sky. This single-template model is similar to the one implemented in \citet{Schlafly_2014}, with the exception that they use the angular distribution of the Planck emission \citep{Planck_2011} instead of CO velocity slices as their dust templates and fix $d_{fore}$ to 0 pc. In comparison to Planck, our more complex multi-slice CO velocity model is better able to trace the underlying molecular $\rm H_2$ and probe the structure of the cloud along the line of sight.

There are a number of limitations to our model, most noticeably that intensity structures in \textit{position-position-velocity (p-p-v)} space do not necessarily correspond to physical density structures in \textit{position-position-position (p-p-p)} space \citep[see][]{Beaumont_2013}. As discussed in \citet{Beaumont_2013}, there are two ways this lack of one-to-one correspondence between density and intensity can manifest itself. First, if there exists an internal velocity gradient, a single density structure could split into two velocity structures. And second, two or more density structures could merge into a single velocity structure, if two structures at two different distances along the line of sight possess the same velocity. With this in mind, we have given our velocity slices the freedom to switch distance order, and also permit placing two different velocity components at the same distance. This addresses the issue of a single density structure splitting into two velocity structures. However, our model is not flexible enough to handle the case where one velocity structure splits into two density structures, at two different distances. If this occurs, the model would be a poor description of the data. By building the ``c" coefficients into our model, controlling how much reddening is assigned to each slice, we should be able to compensate for this to some degree, by assigning adjacent slices more or less reddening. 

We discuss the CO slice distance parameters ($\{d_1,d_2...d_n\}$) in more detail in \S \ref{ds}, the gas-to-dust coefficients ($\{c_1,c_2...c_n\}$) in \S \ref{cs}, the foreground dust cloud parameters in \S \ref{foreground} and our method for handling outliers in \S \ref{outliers}. 

\subsection{Cloud Distance Parameters} \label{ds}
For each cloud of interest (B5, IC345, B1, NGC1333, L1448, and L1451), we have a set of parameters $\{d_1,d_2...d_n\}$ that describe the distance moduli to velocities slices $\{v_1,v_2...v_n\}$ containing the CO emission for that cloud. The typical velocity range of the CO emission observed towards each cloud spans $\rm \approx 5-10 \; km \; s^{-1}$, meaning that, in general, there are several dozen to hundreds of velocity slices along the line of sight. However, because these velocity channels are highly correlated with one another, we choose to downsample the cube along the spectral axis. Because we want to preserve the total CO emission along the line of sight, we downsample by summing to produce downsampled cubes with a velocity slice channel width of $\rm \approx \; 2 \; km \; s^{-1}$. Specifically, towards each region (green polygonal boundaries in Figure \ref{fig:perseus_regions}b), we determine the minimum and maximum velocity slice in which CO emission appears above the noise level \textit{and} the $\rm ^{12}CO$ line is not saturated.\footnote{\label{self_absorbed_footnote} Recall that we determine where $\rm ^{12}CO$ becomes saturated using the analysis of \citet{Pineda_2008}, which finds that $\rm ^{12}CO$ saturates and becomes non-linear with reddening around an $A(V)=4$ mag. Thus, we only consider the CO emission over the same area we select our stars (i.e. sightlines inside the green polygonal boundaries in Figure \ref{fig:perseus_regions}b with corresponding NICER $A(V) < 4$ mag). These areas are shown overlaid with yellow points (constituting our target stars) in Figure  \ref{fig:perseus_regions}c. See discussion in \S \ref{targets} for more details.} We then group the velocity slices across this velocity range into $n$ batches (where $n$ is chosen so that the velocity range of each batch is $\rm \approx 2 \; km \; s^{-1}$), sum the CO emission in each batch of slices, and assign the downsampled slice the \textit{CO intensity-weighted} mean velocity of that batch. The CO intensity-weighted mean slice velocities are computed using only the pixels along sightlines where CO never saturates.\textsuperscript{\ref{self_absorbed_footnote}} This choice of downsampled spectral resolution ($\rm \approx 2 \; km \; s^{-1}$) yields between 3-6 velocity slices towards our target regions, meaning that each cloud is composed of 3-6 distance components ($\{d_1,d_2...d_n\}$). This slice range allows for the freedom to place the velocity slices at multiple distances along the line of sight, while preventing our parameter space from becoming overly redundant or highly covariant. The velocity range of each cloud, along with the number of downsampled velocity slices and the CO intensity-weighted mean velocity of each slice, can be found in Table \ref{velocity_table}.

The average distances we present in \S \ref{results} are robust to our choice of downsampled velocity slice width. We tested our method using two alternatives for the slice width. First, we tested a slice width equal to $\rm \approx 1 \; km \; s^{-1}$. Second, we tested a slice width equal to the entire velocity range for the cloud shown in Table \ref{velocity_table}. In both cases, the average distances agree with those reported in Table \ref{average_distances}. However, as the number of velocity slices increases, the reddening in each slice decreases, and those slices with the least amount of reddening become even less informative to the fit. We need two additional parameters for every velocity slice, and the larger parameter space means the fit takes longer to converge. Thus, while the single velocity slice model is unable to capture the presence of multiple distance components, the choice of a downsampled spectral width $\rm \lesssim 2 \; km \; s^{-1}$ does not provide any additional information about the cloud that cannot be captured with fewer slices.

We place a uniform, flat prior on the distance moduli to the velocity slices \{$d_1$, $d_2$, ...$d_n$\} in the range $\rm 6.5 \; mag < \mu < 8 \; mag$ , corresponding to 200 pc $<$ d $<$ 400 pc. This is consistent with the range of potential distances to the Perseus Molecular Cloud from the literature \citep[see][]{Hirota_2008,Hirota_2011, Cernis_1990, Cernis_1993, Schlafly_2014, de_Zeeuw_1999}. None of our distance parameters, particularly those associated with slices with higher amounts of CO emission, are strongly prior-constrained, so we anticipate our adoption of a flat prior in distance modulus (rather than distance) to have a negligible effect on our results\footnote{This was determined by looking at the corner plots (see the Appendix). If our results were strongly prior constrained, the posteriors for the distance parameters would either be flat over the range $\mu=6.5-8$ mag, or frequently bump up against the boundaries of our flat distance prior. Our corner plots indicate this is not the case for almost all of our distance parameters.}. 

\subsection{Gas-to-Dust Conversion Factor Parameters} \label{cs}
For each of our velocity slices $\{v_1,v_2...v_n\}$ in each cloud, we have a set of parameters $\{c_1,c_2...c_n\}$ that describes how gas relates to dust at each slice distance $\{d_1,d_2...d_n\}$. Specifically, it converts the amount of integrated CO emission in $\rm K \; km \; s^{-1}$ in each downsampled velocity slice to the amount of reddening, E($B-V$), in units of magnitudes. In theory, we could fix this parameter in our model. However, neither of the parameters that constitute this conversion coefficient (i.e. the $X_{CO}$ factor and the $\rm N_{H_2}$-to-reddening factor, as discussed below) is well-constrained, varying in column density and likely physical density of the cloud. We estimate a central value for the prior on the CO-to-reddening coefficient and allow it to vary about this value in our fits. 

To derive an average literature value for this coefficient, we adopt a $\rm ^{12}CO$ $X$-factor of ${\rm 1.8 \times 10^{20} \; cm^{-2} \; K^{-1} \; km^{-1} \; s}$ from \citet{Dame_2001}, where the $X$-factor converts $\rm ^{12}CO$ integrated intensity to $\rm H_2$ column density. Next, assuming that all the hydrogen traced by the reddening is in molecular form---the same assumption made in \citet{Pineda_2008}---we adopt an $\frac{\rm N({H_2})}{E(B-V)}$ ratio of $\rm 2.9 \times 10^{21} \; cm^{-2} \; mag^{-1}$ from \citet{Bohlin_1978}, which converts $\rm H_2$ column density to reddening. Combining the two factors, we get an average $\frac{E(B-V)}{W(CO)}$ factor equal to $0.062 \rm \; \frac{mag}{K \; km \; s^{-1}} $. In practice, we integrate over the $\rm \approx \; 2 \; km \; s^{-1}$ velocity channels (defined in \S \ref{ds}) by summing the CO emission in Kelvin and multiplying by the native channel width ($\rm 0.064 \; km \; s^{-1}$) of the CO COMPLETE survey. This gives the CO brightness in each velocity slice in $\rm K \; km \; s^{-1}$, which we can then multiply by the $\frac{E(B-V)}{W(CO)}$ factor sampled at each iteration to produce the total E($B-V$) in magnitudes.

We place a log-normal prior on our $\frac{E(B-V)}{W(CO)}$ factors, \{$c_1$, $c_2$, ...$c_n$\}, which ensure this factor is always positive. For simplicity, we treat the \{$c_1$, $c_2$, ...$c_n$\} parameters as multiplicative factors in front of the average gas-to-dust coefficient we adopt from the literature ($0.062 \rm \; \frac{mag}{K \; km \; s^{-1}} $, as computed above). A value of $c=1$ would yield a gas-to-dust conversion coefficient of $0.062 \rm \; \frac{mag}{K \; km \; s^{-1}} $, while a value of $c=2$ would yield a gas-to-dust conversion coefficient of $0.124 \rm \; \frac{mag}{K \; km \; s^{-1}} $. We set the mean of the prior equal to the average value from the literature ($\bar{c}=1$) and set the standard deviation to $\sigma=0.2$, where $\sigma$ is the standard deviation of the natural logarithm of the $c$ parameters. The larger value of $\sigma$ constitutes a relatively loose prior on the conversion coefficients, and is consistent with the underlying uncertainty in this factor from the literature.\footnote{It is well known that the formation of CO (and thus the $X$-factor) is dependent on the metallicity of the gas and the strength of the background UV radiation field, which can cause CO molecules to disassociate \citep{Shetty_2011, Pineda_2008}. \citet{Pineda_2008} finds that the $X_{CO}$ factor within Perseus can vary by at least a factor of two, depending on which star-forming region is used in the fit (e.g. NGC1333 vs. B5; see their Table 4) or whether it is determined in saturated or unsaturated regimes.} As discussed further in \S \ref{results}, we tend to infer $c$ coefficients in the range $0.022-0.070 \rm \frac{mag}{K \; km \; s^{-1}} $, which is lower on average than the mean we adopt for our Gaussian prior---indicating that this prior is being ``tugged" on to produce a smaller value for the gas-to-dust coefficient. This is consistent with the findings of \citet{Pineda_2008} (see \S \ref{results} for more details).

\subsection{Foreground Dust Cloud Parameters} \label{foreground}
Our model includes two foreground dust cloud parameters, whose reddening contribution is completely independent from the CO emission associated with the Perseus Molecular Cloud. Specifically, we parameterize the foreground dust cloud as lying at a distance modulus $d_{fore}$, and with a reddening contribution in $\Delta {\rm E}_{B-V}$ given by $E_{fore}$. We assume there are no background clouds beyond the distance to Perseus, which is consistent with the bulk of the reddening profiles given by the \citetalias{Green_2018} method for sightlines towards our regions of interest. Introducing the set of foreground dust cloud parameters \{$d_{fore}$, $E_{fore}$\} serves two purposes. 

First, towards every region, we observe a set of stars that lie at distances prior to Perseus ($d \lesssim 200$ pc) and whose reddening lies between zero and the bulk of the reddening associated with Perseus' CO features. Thus, there is a physical basis for including these parameters. The second reason is that the foreground dust cloud accommodates the fact that $\rm ^{12}CO$ is destroyed by the interstellar radiation field until $A(V) \lesssim 1$ mag, at which point it becomes self-shielded. The $E_{fore}$ parameter then accounts for the reddening in a regime where CO and reddening are non-linear, and corrects for the ``missing" reddening below $A(V) \approx 1.0$ mag which appears in the stellar posteriors but not in the total line-of-sight CO emission. As a result, this parameter is somewhat correlated with the CO self-shielding threshold. Since we do not take into account the critical density of CO, the model will not be a perfect description of the data. However, we are most interested in the distances to the velocity slices (\S \ref{ds}), which are only very weakly covariant with our nuisance parameters (i.e. the $d_{fore}$ and $E_{fore}$ parameters described in this section, plus the gas-to-dust coefficients described in \S \ref{cs}). 

Like the cloud distances, we place a flat prior on the distance modulus to the foreground reddening cloud $d_{fore}$, where the lower bound is the minimum distance modulus represented in our gridded stellar posteriors ($\mu=4$) and the upper bound is the minimum distance modulus to the velocity slices corresponding to Perseus, \{$d_1$, $d_2$, ...$d_n$\}. We additionally place a flat prior on the reddening of the foreground dust cloud, $E_{fore}$, restricting it to between 0 and no greater than the average CO-based reddening to stars towards the cloud. The average reddening is computed by taking the sum of the each slice's CO emission times its gas-to-dust coefficient for all stars towards the cloud, and then taking the median of these values. 

\subsection{Accounting for Outliers} \label{outliers}
\begin{figure}[h!]
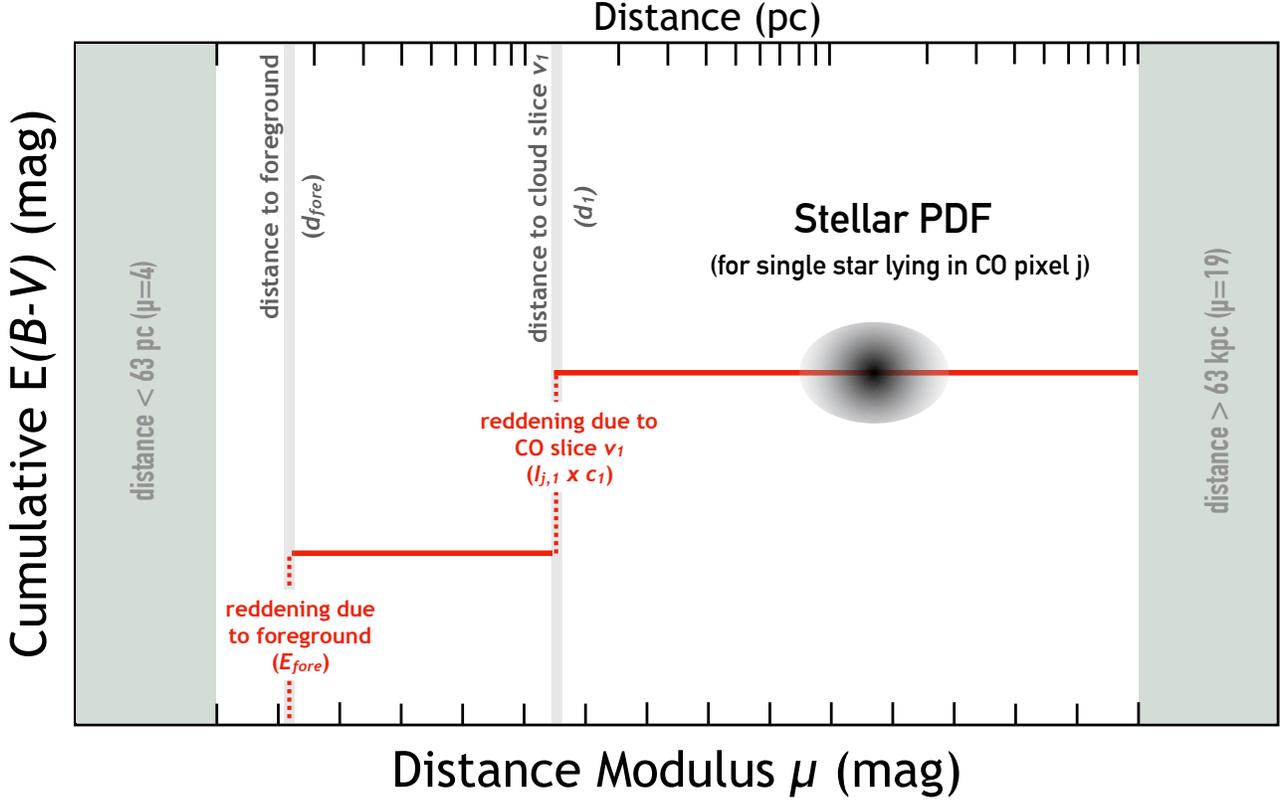

\begin{center}
\includegraphics[width=1.0\columnwidth]{{{profile_schematic}}}
\caption{{\label{fig:example_profile} Cartoon line-of-sight reddening profile assuming a model with a single velocity slice $v_1$ (where the cloud can be described by single distance component $d_1$). Such a reddening profile is defined by the free parameters $d_{fore}$ (foreground cloud distance), $E_{fore}$ (foreground cloud reddening), $d_1$ (distance modulus to single velocity slice) and $c_1$ (gas-to-dust conversion coefficient in single velocity slice). The quantity $I_{j,1}$ constitutes the CO emission for star $i$ lying in pixel $j$ in the single velocity slice $v_1$.  The reddening profile is overlaid on an idealized joint posterior on distance and reddening (grayscale ellipsoid) for individual star $i$ lying in CO pixel $j$. The star's likelihood is simply the integral over $\mu$ following the reddening profile (solid red line); the dashed red lines are excluded from the integration. In \S \ref{MCMC} the likelihood from this star would be multiplied together with the likelihoods of thousands of other stars to get the total likelihood contribution (see Equation \ref{pdf_function}).}}
\end{center}
\end{figure}

We include one final free parameter in our model---denoted $P_b$---to reduce the influence of outlier stars on our Equation \ref{pdf_function}. In the post-Gaia era this is important for a small, but non-trivial number of stars, given that our gridded stellar posteriors can sometimes be incorrect due to errors in the stellar photometry, the adoption of a fixed extinction curve, or the presence of stars not well-accounted for in our models (e.g. white dwarfs, young stellar objects, quasars, variable stars). While the distance uncertainty for stars with reliable Gaia DR2 parallax measurements is mitigated, Gaia will have no effect on the reddening we infer for the star. To this purpose, a final free parameter $P_b$ is implemented as part of the Gaussian mixture model described in \citet{Hogg_2010}. $P_b$ quantifies the probability that any individual star is ``bad" and unlikely to be drawn from our model. Following Eqn. (17) in \citet{Hogg_2010}, the total likelihood for an individual star $i$ in the context of our mixture model is given by:

\begin{equation}
L_{\star,i}=P_b\times L_{bad, i} + (1-P_b)\times L_{good, i}.  
\end{equation}
 
\noindent where $L_{good,i }$ is the original likelihood for star $i$ (computed by taking an integral following ${\rm E}_{B-V}(\mu_i; \vec{\alpha})$ over $\mu$ through the star's distance and reddening posterior).\footnote{This integral is the same as given in Equation \ref{pdf_function}: $\int p(\mu_i, E_{B-V}(\mu_i; \vec{\alpha}) \mid \vec{m_i}) \; d\mu_i$ and is equivalent to summing over the solid red line in Figure \ref{fig:example_profile}.} $L_{bad,i }$ is the same integral taken through a stellar posterior array that is flat in reddening, but not in distance (i.e. the probability is localized to one column of reddening in the stellar posterior array). This is similar to adding a small constant to the likelihood of every star, and is well-suited to our modeling, given that a majority of the stars in our sample have distances informed by Gaia.

We impose a truncated log-normal prior on our $P_b$ parameter, with a mean of 0.05 and a $1\sigma$ range of [0.025, 0.10], such that $0 < P_b < 1$. This is based on prior knowledge that $\approx 90\%$ of stars lie on the main sequence in the solar neighborhood \citep{Bahcall_1980} and should be accurately captured by the stellar models used to derive our per-star posteriors on distance and reddening from \citetalias{Green_2018}.

\section{Parameter Estimation using Nested Sampling} \label{MCMC}
For each of our target regions (see Figure \ref{fig:perseus_regions}) we sample a set of $3+2n$ model parameters ($d_{fore}$, $E_{fore}$, $P_b,$ \{$d_1$, $d_2$, ...$d_n$\}, \{$c_1$, $c_2$,...$c_n$\}, where $n$ is the number of velocity slices) using the nested sampling code \href{https://dynesty.readthedocs.io/en/latest/}{\texttt{dynesty}}\footnote{\url{https://dynesty.readthedocs.io}} (Speagle et al. 2018, in prep). Nested sampling \citep{Skilling_2006} is similar to traditional MCMC algorithms in that it generates samples that can be used to estimate the posterior PDF given in Equation \ref{pdf_function}. The nested sampling algorithm relies on iteratively drawing samples (or ``live" points) from the constrained prior distribution, where the likelihood value of a new sample must be greater than the lowest likelihood value of existing samples. In this way, the live point with the lowest likelihood is replaced by a new live point of higher likelihood at every iteration. As this process progresses, the live points sample a smaller and smaller region of the prior ``volume". One continues sampling until some stopping criterion is reached and the remaining live points occupy the region of highest likelihood. 

There are several reasons to use \texttt{dynesty} over more commonly used affine-invariant ensemble MCMC samplers. Ensemble MCMC sampling codes like \texttt{emcee} \citep{Foreman_Mackey_2013} have been shown to have ``undesirable properties" in higher-dimensional parameter spaces ($n_{param} \gtrsim 10-20$). When the number of model parameters is high, ensemble samplers may not only fail to converge to the target distribution, but also visually appear converged without have done so \citep{Huijser_2015}. Nested sampling has been shown to perform well in higher-dimensional parameter spaces and is also efficient at exploring multi-modal distributions \citep{Feroz_2009, Handley_2015}. Additional benefits of \texttt{dynesty} include a user-defined stopping criterion that can act as a convergence metric, so the number of generated samples does not have to be pre-defined beforehand. Finally, \texttt{dynesty} can quantify the statistical uncertainties in a single run using a combined reweighting/bootstrapping procedure \citep{Higson_2017b, Higson_2017a}.

Since \texttt{dynesty} can be run in multiple modes, we provide the exact setup used to derive our results in \S \ref{dynesty_appendix} of the Appendix. While less efficient at producing independent samples, using a more traditional sampler from \texttt{emcee} produces results that are consistent with the dynamic nested sampler when comparing runs for the cloud with the largest parameter space (NGC1333).\footnote{In more detail, in order to test whether our results are robust to changes in sampler, we repeat the analysis for NGC1333 using the affine-invariant MCMC ensemble sampler from \texttt{emcee} \citep{Foreman_Mackey_2013}. Specifically, we run with 100 walkers, where each walker takes 20,000 steps. We set $thin=10$ so only every tenth sample is saved. Prior to this, we run 1000 burn-in steps. Using the stacked chain (flattened along the walker axis), we compute the median of the samples from the chain for each parameter. We find that the median of the samples agrees with our \texttt{dynesty} results within the uncertainties we report in Table \ref{results_table}. Since we report the 16th and 84th percentiles of the \texttt{dynesty} samples via the upper and lower bounds, this means that the 50th percentile of the samples derived from our \texttt{emcee} run falls within the 16th and 84th percentiles of the samples from our \texttt{dynesty} run.}

\begin{center} 
\begin{figure}
   \centering
    \includemedia[
    activate=onclick,
    flashvars={
        modestbranding=1 
        &autohide=1 
        &showinfo=0 
        &rel=0 
    }
            ]{\includegraphics[width=1.0\linewidth]{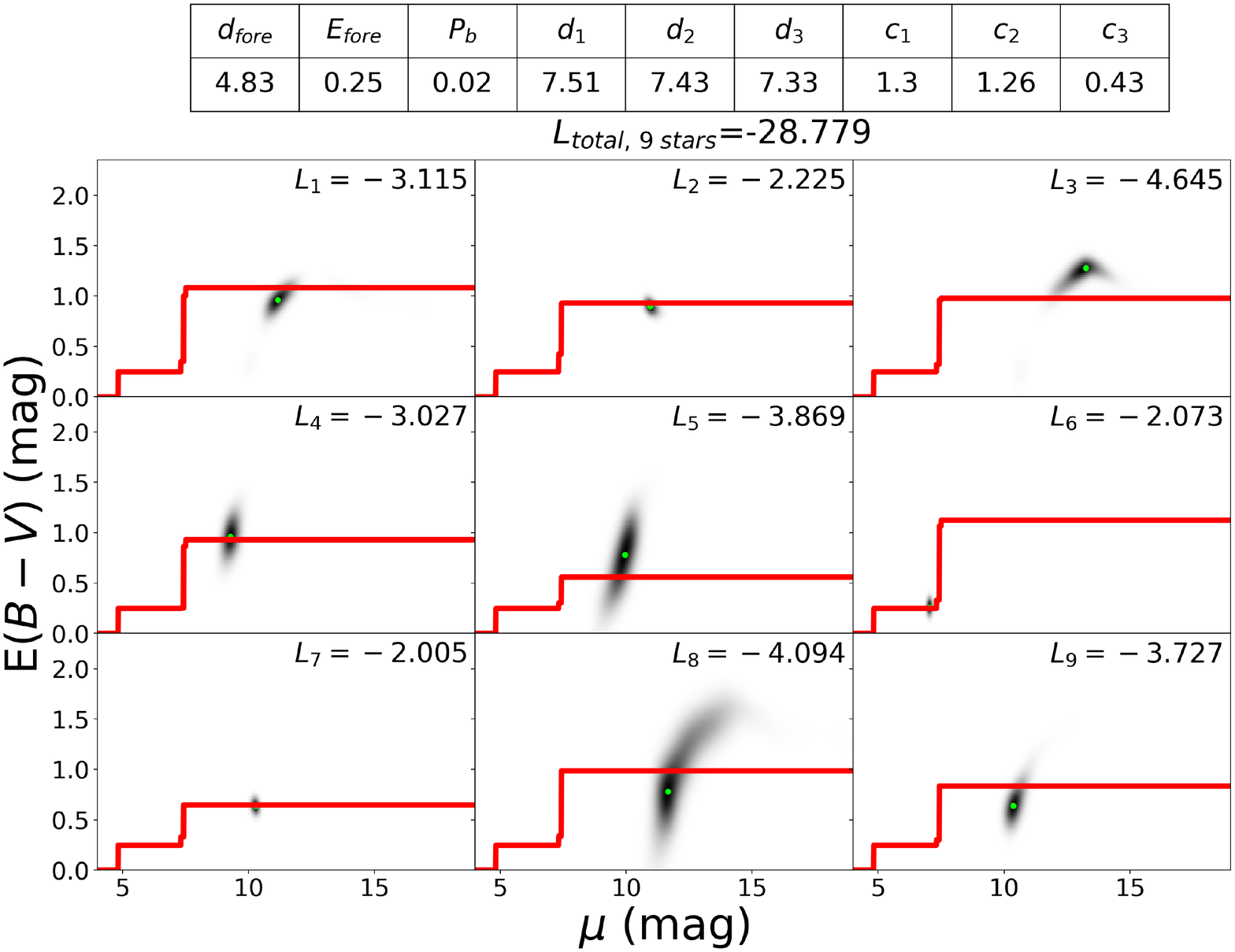}}{https://youtube/v/t_iWrkB8dFQ}
\caption{\label{fig:video} If this video is not supported in your document viewer, click \href{https://youtu.be/t_iWrkB8dFQ}{here}. Video showing samples (summarized in the table at top) from our \texttt{dynesty} run towards the B5 region. \textit{These are not fair samples, but are intended to illustrate how variations in the reddening profile affect the log-likelihood}. Nine stars (out of thousands) are selected towards the region. Each star lies in a different CO pixel, so the reddening profile varies from star to star and is dependent on the CO emission in each velocity slice corresponding to that pixel. The reddening profile is built using the sample summarized in the table at top. Individual stellar log-likelihoods (listed in the upper right corner of each panel) are calculated by integrating the reddening profile (red lines) over $\mu$ through each star's distance-reddening posterior (background grayscale of each panel) and taking the logarithm. In each stellar posterior, the region of highest probability is marked with a lime green scatter point. The total log-likelihood for all nine stars ($L_{tot, \, 9 \, stars}$) is listed above the panels.}
\end{figure}
\end{center}

To get a sense of how the sampling operates, in Figure \ref{fig:video} we show a video which illustrates the progression of our \texttt{dynesty} samples over the course of a run. We select nine stars used in our B5 fit. For each star, we use the current sample (summarized in the table at top) along with the CO emission coincident with each star to build up the star's reddening profile (red lines), which we then overlay on its distance-reddening posterior from \citetalias{Green_2018} (grayscale background in each panel). Integrating over $\mu$ along each reddening profile, we show the individual log-likelihood for each star in the upper right hand corner of every panel. Adding all these individual log-likelihoods together, we get the total-log likelihood for this batch of nine stars ($L_{tot, \, 9 \; stars}$), which we list atop the panels. The samples are drawn sequentially from the \texttt{dynesty} chain as we converge towards the region of highest likelihood. However, the total log-likelihood for these nine stars does not necessarily increase as the video progresses, since the samples are determined by the best fit to the full sample of stars rather than just these nine. 

Finally, we note that when calculating the line-of-sight integral for the individual log-likelihoods, we interpolate between cells in the two-dimensional stellar posterior array. This mitigates the negative effects of binning on our results. 

\section{Results} \label{results}
In Table \ref{results_table}, we summarize the results for the six star-forming regions targeted in this study---L1451, L1448, NGC1333, B1, IC348, and B5. Specifically, we report the 50th percentile of the samples for each parameter: $d_{fore}$, $E_{fore}$, $P_b,$ \{$d_1$, $d_2$, ...$d_n$\}, \{$c_1$, $c_2$,...$c_n$\}, where again, $n$ is determined by the number of downsampled velocity slices towards each cloud (see Table \ref{velocity_table}). In order to properly estimate the posterior, we weight the samples by their posterior mass, as discussed in the \href{http://dynesty.readthedocs.io/en/latest/overview.html?highlight=weight}{\texttt{dynesty} documentation}. For the \{$c_1$, $c_2$,...$c_n$\} parameters, recall that we report the value of a multiplicative factor in front of the nominal gas-to-dust coefficient we adopt from the literature (see \S \ref{cs}). In addition to reporting the median value, we provide the 16th and 84th percentiles via the upper and lower bounds, equivalent to the $1\sigma$ range for a Gaussian distribution. We also include corner plots---showing different projections of our \texttt{dynesty} samples in the $2n+3$-dimensional parameter space---in the Appendix. 

We generally favor a smaller CO-to-reddening conversion factor than the nominal factor we derive from the literature (\S \ref{cs}), suggesting a smaller $X_{CO}$ factor than given in \citet{Dame_2001} or a larger $\rm N_{H_2}$-to-reddening factor than given in \citet{Bohlin_1978}. Our mean gas-to-dust coefficient across all regions is $0.72\times c$, where c is the nominal coefficient we adopt from the literature, equal to $0.062 \rm \; \frac{mag}{K \; km \; s^{-1}}$. The argument for a smaller $X_{CO}$ factor is consistent with the results of \citet{Pineda_2008}. After accounting for the CO self-shielding threshold, they find that the $X_{CO}$ in unsaturated regions is ${\rm 0.72 \times 10^{20} \; cm^{-2} \; K^{-1} \; km^{-1} \; s}$. Assuming that our $\rm N_{H_2}$-to-reddening factor is accurate, our typical $c$ coefficients results are consistent with an $X_{CO}$ factor of  ${\rm 1.30 \times 10^{20} \; cm^{-2} \; K^{-1} \; km^{-1} \; s}$, which is lower than our adopted \citet{Dame_2001} $X_{CO}$ factor (${\rm 1.8 \times 10^{20} \; cm^{-2} \; K^{-1} \; km^{-1} \; s}$), but not quite as low as the \citet{Pineda_2008} value. 

\begin{figure}[h!]
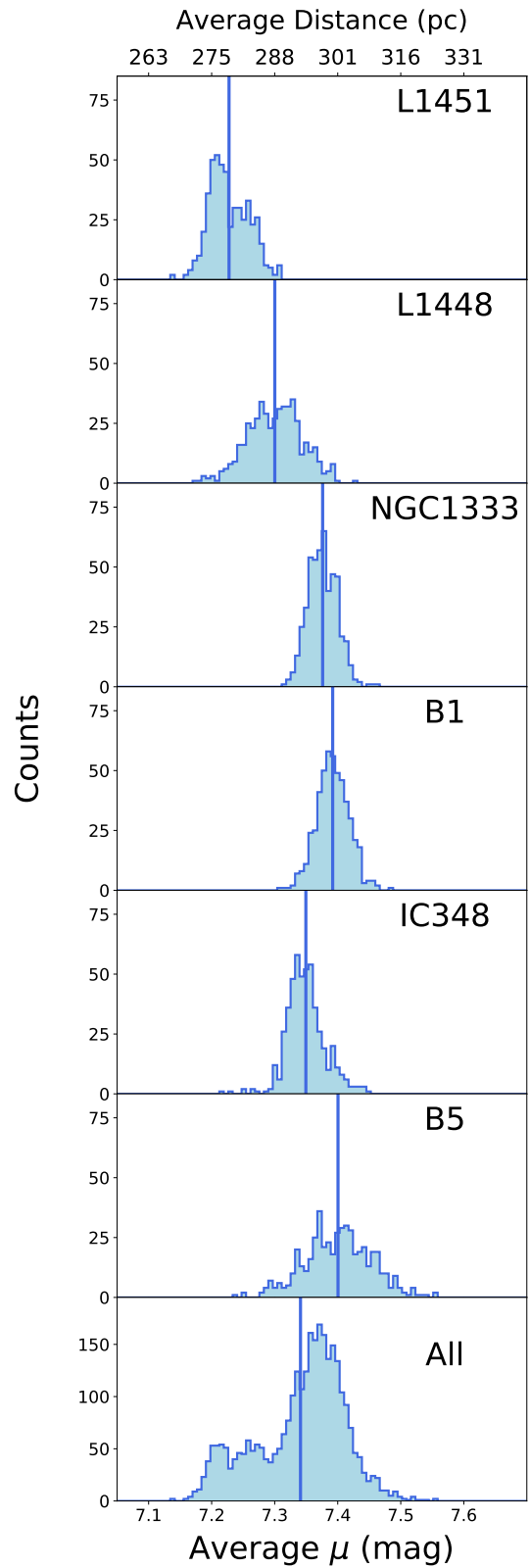

\begin{center}
\includegraphics[width=0.4\columnwidth]{{{hist_outlier_finalpub_oct1}}}
\caption{{\label{fig:hist_dist_comp} Monte Carlo realizations of the average reddening-weighted distance to each cloud as described at the beginning of \S \ref{results}. Each ``count" is a reddening-weighted average distance computed using Equation \ref{avg_dist_eqn}. From top to bottom, the panels are sorted in order of increasing Right Ascension. We report the mean of the distribution in each panel (vertical blue lines) as the final overall average distance to each star-forming region in Table \ref{average_distances}. Likewise, the uncertainty we report in Table \ref{average_distances} is the standard deviation of each distribution. An average distance and distance uncertainty for the Perseus complex as a whole (bottom panel) is computed by stacking the average distances to individual regions in the first six panels.}}
\end{center}
\end{figure}

We discuss the properties of individual clouds in comparison with the literature values in more detail in the next section. We also provide an average distance and distance uncertainty to each cloud (Table \ref{average_distances}). The average distances are computed using the distances to the set of velocity slices towards each cloud, and we weight each slice according to its total contribution to the line-of-sight reddening. Specifically, to compute the average distance to each cloud we perform a Monte Carlo-based averaging procedure where samples are drawn at random (again, weighted by their posterior mass) from ``noisy" realizations of our original \texttt{dynesty} chain. For each Monte Carlo draw we first construct the noisy set of samples using \texttt{dynesty's} \href{http://dynesty.readthedocs.io/en/latest/errors.html?highlight=simulate\_run\#combined-uncertainties}{simulate\_run()} function. This process uses a combination of bootstrapping and jittering to add random noise to the original samples. This allows us to account for statistical uncertainties in the average cloud distances we report in Table \ref{average_distances}. Then, for each realization, after drawing from our set of noisy samples, we compute a reddening-weighted average distance to the cloud using the following formula: 

\begin{equation} \label{avg_dist_eqn}
\langle d \rangle=\frac{\sum\limits_k^{n_{slices}} \langle I_{CO} \rangle_k \; c_k \; d_k}{\sum\limits_k^{n_{slices}} \langle I_{CO} \rangle_k \; c_k}
\end{equation}

\noindent where $\langle I_{CO} \rangle_k$ is the mean CO emission in the $kth$ velocity slice, $c_k$ is the gas-to-dust coefficient in the $kth$ slice for each realization, and $d_k$ is the distance to the $kth$ slice for each realization. We repeat this process 500 times, producing 500 realizations of the average reddening-weighted distance to the cloud. The final average distance and uncertainty to each cloud that we report in Table \ref{average_distances} are the mean and standard deviation of the set of 500 average reddening-weighted distances. We show histograms of the Monte Carlo realizations of the average reddening-weighted distances towards each region in Figure \ref{fig:hist_dist_comp}, which are discussed further in \S \ref{discussion}. We compute an average distance for the entire complex as a whole by stacking the averages from all regions, and taking the mean and standard deviation. This is shown in the bottom panel of Figure \ref{fig:hist_dist_comp}. We compute an average distance to the entire Perseus complex of 294 pc, with a statistical uncertainty of 10 pc.

In Table \ref{average_distances} we also report the average reddening-weighted velocity for each cloud. Since the CO intensity-weighted mean velocity of each slice is fixed (see Table \ref{velocity_table} and \S \ref{ds} for details on how this is calculated) and our $c$ coefficients are well-constrained, we do not perform a Monte Carlo averaging procedure, but simply weight each velocity slice by the mean CO in each slice times the median $c$ coefficient for that slice (summarized in Table \ref{results_table}). We also report the peak reddening velocity for each cloud; this corresponds to the velocity of the slice with the highest reddening, and which dominates the total dust column density along the line of sight (see Table \ref{velocity_table} for the velocities of the slices we consider towards each region).

Finally, we emphasize that in addition to the statistical uncertainty reported in Table \ref{average_distances}, there is also systematic uncertainty that needs to be taken into account. \citet{Schlafly_2014} quantified the systematic uncertainty to be at the $\approx 10\%$ level due to the reliability of our stellar models and the adoption of a fixed extinction curve, both of which should affect the shape of our distance-reddening posteriors. However, with the inclusion of the Gaia DR2 data, these effects should be significantly reduced, particularly because stars foreground to the cloud ($\mu \lesssim 8$ mag) have very well constrained distances, with \texttt{parallax\_over\_error} values of $5:1$ on average; these stars carry significant weight when fitting the distances to the velocity slices. Nevertheless, there are ancillary choices we have made, independent from our stellar models and fixed extinction curve, that will also have some small effect on our results. This includes our choice of grid (e.g. bin size) for our distance-reddening posteriors, how much we smooth along the reddening axis (see \S \ref{rv}), our choice of outlier model (see \S \ref{outliers}), or how we filter stars based on their best-fit chi-squared values (see \S \ref{targets}). We have rerun our line-of-sight fits for various combinations of these choices and found that the results reported in Table \ref{average_distances} are robust to $0.1$ mag in distance modulus, or $\approx 5\%$ in distance. As a result, we conservatively recommend that a systematic uncertainty of 0.1 mag in distance modulus ($\approx 5 \%$  in distance) be adopted in quadrature with our statistical uncertainties.  In Table \ref{average_distances}, the first uncertainty term reported is the statistical uncertainty, while the second is the systematic uncertainty.

\begin{turnpage}
\renewcommand{\arraystretch}{2} 
\begin{deluxetable}{cccccccccccccccc}
\colnumbers
\tablecaption{\label{results_table}}
\tabletypesize{\scriptsize}
\colnumbers
\setlength{\tabcolsep}{2pt}
\tablehead{[-.3in] \colhead{Cloud} & \colhead{$d_{fore}$} & \colhead{$E_{fore}$} & \colhead{$P_b$} & \colhead{$d_1$} & \colhead{$d_2$} & \colhead{$d_3$} & \colhead{$d_4$} & \colhead{$d_5$} & \colhead{$d_6$} & \colhead{$c_1$} & \colhead{$c_2$} & \colhead{$c_3$}  & \colhead{$c_4$} & \colhead{$c_5$} & \colhead{$c_6$} \\[-.1in] \colhead{} & \colhead{mag} & \colhead{mag} & \colhead{} & \colhead{mag} & \colhead{mag} & \colhead{mag} &  \colhead{mag} & \colhead{mag} & \colhead{mag} & \colhead{} & \colhead{} & \colhead{} & \colhead{} & \colhead{} & \colhead{} \\[-.2 in] \colhead{} & \colhead{} & \colhead{} & \colhead{} & \colhead{pc} & \colhead{pc} & \colhead{pc} &  \colhead{pc} & \colhead{pc} & \colhead{pc} & \colhead{} & \colhead{} & \colhead{} & \colhead{} & \colhead{} & \colhead{}}
\startdata
\textbf{B5} & $4.50^{+0.39}_{-0.32}$ & $0.25^{+0.01}_{-0.01}$ & $0.02^{+0.00}_{-0.00}$* & $7.56^{+0.12}_{-0.11}$ & $7.40^{+0.05}_{-0.06}$ & $7.34^{+0.10}_{-0.08}$ & &  &  & $1.34^{+0.08}_{-0.08}$ & $1.24^{+0.02}_{-0.03}$ & $0.47^{+0.03}_{-0.03}$ &  &  & \\ [-.1in] & $79^{+16}_{-13}$ &  &  & $325^{+18}_{-17}$ & $302^{+8}_{-9}$ & $294^{+13}_{-11}$ & &  &  &  &  &  &  &  & \\ [.25in]
\textbf{IC348} & $6.14^{+0.03}_{-0.04}$ & $0.35^{+0.00}_{-0.00}$* & $0.03^{+0.00}_{-0.00}$* & $7.35^{+0.05}_{-0.04}$ & $7.35^{+0.04}_{-0.03}$ & $7.34^{+0.04}_{-0.03}$ & $7.35^{+0.05}_{-0.03}$ &  &  & $1.18^{+0.05}_{-0.05}$ & $1.04^{+0.03}_{-0.03}$ & $1.08^{+0.01}_{-0.01}$ & $0.43^{+0.03}_{-0.03}$ &  & \\[-0.1in] & $169^{+2}_{-3}$ &  &  & $295^{+7}_{-5}$ & $295^{+5}_{-4}$ & $294^{+5}_{-4}$ & $295^{+7}_{-4}$  &  &  &  &  &  &  &  & \\[0.25in]
\textbf{B1} & $4.45^{+0.67}_{-0.34}$ & $0.15^{+0.01}_{-0.01}$ & $0.03^{+0.00}_{-0.00}$ & $7.13^{+0.22}_{-0.25}$ & $7.36^{+0.16}_{-0.23}$ & $7.48^{+0.04}_{-0.03}$ & $7.38^{+0.03}_{-0.03}$ & $7.45^{+0.04}_{-0.06}$ &  & $0.63^{+0.05}_{-0.05}$ & $0.38^{+0.03}_{-0.03}$ & $0.41^{+0.02}_{-0.02}$ & $0.97^{+0.02}_{-0.02}$ & $0.45^{+0.03}_{-0.03}$ & \\[-0.1in] & $78^{+28}_{-13}$  &  &  & $266^{+28}_{-32}$ & $297^{+23}_{-33}$ & $313^{+6}_{-5}$ & $300^{+4}_{-4}$   & $309^{+5}_{-8}$   &  &  &  &  &  &  & \\[0.25in]
\textbf{NGC1333} & $5.20^{+0.43}_{-0.85}$ & $0.10^{+0.01}_{-0.01}$ & $0.02^{+0.00}_{-0.00}$* & $7.08^{+0.70}_{-0.34}$ & $7.40^{+0.04}_{-0.04}$ & $7.29^{+0.15}_{-0.10}$ & $7.37^{+0.06}_{-0.10}$ & $7.39^{+0.03}_{-0.02}$ & $7.42^{+0.05}_{-0.05}$ & $0.33^{+0.03}_{-0.04}$ & $0.97^{+0.03}_{-0.03}$ & $0.77^{+0.03}_{-0.03}$ & $0.44^{+0.02}_{-0.02}$ & $0.83^{+0.01}_{-0.01}$ & $0.32^{+0.02}_{-0.02}$\\[-0.1in] & $110^{+24}_{-52}$  &  &  & $260^{+99}_{-44}$ & $302^{+5}_{-6}$ & $287^{+20}_{-13}$ & $298^{+9}_{-14}$  & $301^{+4}_{-3}$  & $305^{+7}_{-7}$  &  &  &  &  &  & \\[0.25in]
\textbf{L1448} & $5.64^{+0.25}_{-0.56}$ & $0.16^{+0.00}_{-0.01}$ & $0.02^{+0.00}_{-0.00}$* & $7.30^{+0.11}_{-0.12}$ & $7.30^{+0.08}_{-0.08}$ & $7.34^{+0.08}_{-0.08}$ & $7.28^{+0.07}_{-0.06}$ & $7.34^{+0.07}_{-0.08}$ &  & $0.90^{+0.07}_{-0.07}$ & $0.36^{+0.02}_{-0.02}$ & $0.37^{+0.03}_{-0.03}$ & $0.91^{+0.02}_{-0.02}$ & $1.15^{+0.05}_{-0.04}$ & \\[-0.1in] & $134^{+16}_{-40}$  &  &  & $288^{+15}_{-17}$ & $289^{+10}_{-11}$ & $294^{+11}_{-12}$ & $286^{+9}_{-8}$   & $293^{+10}_{-11}$   &  &  &  &  &  &  & \\[0.25in]
\textbf{L1451} & $4.09^{+0.10}_{-0.06}$ & $0.07^{+0.01}_{-0.01}$ & $0.02^{+0.00}_{-0.00}$* & $7.24^{+0.15}_{-0.17}$ & $6.74^{+0.10}_{-0.11}$ & $7.33^{+0.04}_{-0.03}$ & $6.72^{+0.20}_{-0.15}$ &  &  & $0.53^{+0.03}_{-0.04}$ & $0.32^{+0.02}_{-0.02}$ & $1.30^{+0.04}_{-0.03}$ & $0.30^{+0.03}_{-0.03}$ &  & \\[-0.1in] & $66^{+3}_{-2}$ &  &  & $280^{+20}_{-22}$ & $223^{+11}_{-11}$ & $292^{+6}_{-4}$ & $220^{+21}_{-16}$  &  &  &  &  &  &  &  & \\[0.25in]
\enddata
\tablenotetext{*}{The upper and lower bounds on these parameters are uncertain to $< 0.005$ and have been rounded off to two significant figures}
\tablecomments{The results of our parameter estimation for major star-forming regions across the Perseus Molecular Cloud. For each parameter, we report the 50th percentile of the samples. The 16th and 84th percentiles are given via the upper and lower bounds. The columns are summarized as follows: \textbf{(1)} Name of the star-forming region. \textbf{(2)} the distance modulus of the foreground reddening cloud. Directly under it we list the corresponding distance in parsecs. \textbf{(3)} the reddening in ${\rm E}(B-V)$ of the foreground cloud. \textbf{(4)} the fraction of ``bad" stars, implemented as part of a Gaussian mixture model in an attempt to mitigate outliers.  \textbf{(5)-(10)} the distance moduli of the CO velocity slices corresponding to each region (see Table \ref{velocity_table}). Directly under them we list the corresponding distances in parsecs. \textbf{(11)-(16)} multiplicative factors in front of the nominal gas-to-dust coefficient adopted in this work ($\bar{c}=\rm 0.062 \; \frac{mag}{K \; km \; s^{-1}}$), which are built into the model to account for the uncertainty in how we translate CO emission to reddening.}
\end{deluxetable}
\end{turnpage}

\begin{deluxetable}{ccccc}
\colnumbers
\tablecaption{ \label{average_distances}}
\tabletypesize{\small}
\setlength{\tabcolsep}{8.0pt}
\tablehead{\colhead{Cloud Name} & \colhead{Average Distance Modulus} & \colhead{Average Distance}  & \colhead{Average Velocity} & \colhead{Peak Reddening Velocity}\\ \colhead{} & \colhead{mag} & \colhead{pc} & \colhead{$\rm km \; s^{-1}$} & \colhead{$\rm km \; s^{-1}$}}
\startdata
L1451 & $7.23 \pm 0.03 \pm 0.10$ & $279 \pm 4 \pm 13$ & 3.9 & 4.2 \\
L1448 & $7.30 \pm 0.04 \pm 0.10$ & $288 \pm 6 \pm 13$ & 4.2 & 4.8 \\
NGC1333 & $7.38 \pm 0.02 \pm 0.10$ & $299 \pm 3 \pm 14$ & 5.7 & 7.3 \\
B1 & $7.39 \pm 0.03 \pm 0.10$ & $301 \pm 4 \pm 14$ & 6.2 & 6.7 \\
IC348 & $7.35 \pm 0.03 \pm 0.10$ & $295 \pm 4 \pm 14$ & 8.5 & 9.0 \\
B5 & $7.40 \pm 0.05 \pm 0.10$ & $302 \pm 7 \pm 14$ & 9.7 & 9.7 \\
\hline
Entire Complex & $7.34 \pm 0.07 \pm 0.10$ & $294 \pm 10 \pm 14$ &  &  \\
\enddata
\tablecomments{Average region-by-region distances and velocities for clouds across the Perseus complex. In column \textbf{(1)}, we list the name of the star-forming region. In column \textbf{(2)}, we show the average reddening-weighted distance to the downsampled velocity slices towards each region (see the beginning of \S \ref{results} for details on how this is calculated). The first uncertainty term is the statistical uncertainty, while the second is the systematic uncertainty (estimated to be $\mu=0.1$ mag or  $\approx 5\%$ in distance; see \S \ref{results}). In column \textbf{(3)}, we convert the average distance modulus from column (2) into its corresponding distance in parsecs. As in column (2), the first uncertainty term is the statistical uncertainty, while the second is the systematic uncertainty (estimated to be $\mu=0.1$ mag or $\approx 5\%$ in distance; see \S \ref{results}). In column \textbf{(4)}, we show the reddening-weighted average velocity for each cloud (see \S \ref{results}). In column \textbf{(5)} we list the velocity slice which contains the highest amount of reddening, dominating the dust column density along the line of sight (see column (5) in Table \ref{velocity_table} for the velocity slices we consider towards each region).}
\end{deluxetable}

\subsection{B5}
We consider three velocity slices towards B5 (with CO intensity-weighted velocities of 8.3, 9.7, and $11.0 \; \rm km \; s^{-1}$) and determine $\mu$ of $7.56^{+0.12}_{-0.11}$, $7.40^{+0.05}_{-0.06}$, and $7.34^{+0.10}_{-0.08}$ mag, respectively. This corresponds to distances of $325^{+18}_{-17}$, $302^{+8}_{-9}$, and $294^{+13}_{-11}$ pc. All slices are consistent with being at the same distance. We find an average distance to B5 of $\mu = 7.40 \pm 0.05$ mag ($302 \pm 7$ pc). The difference in distance between L1451 and B5 is roughly 25 pc. 

\begin{figure}[h!]
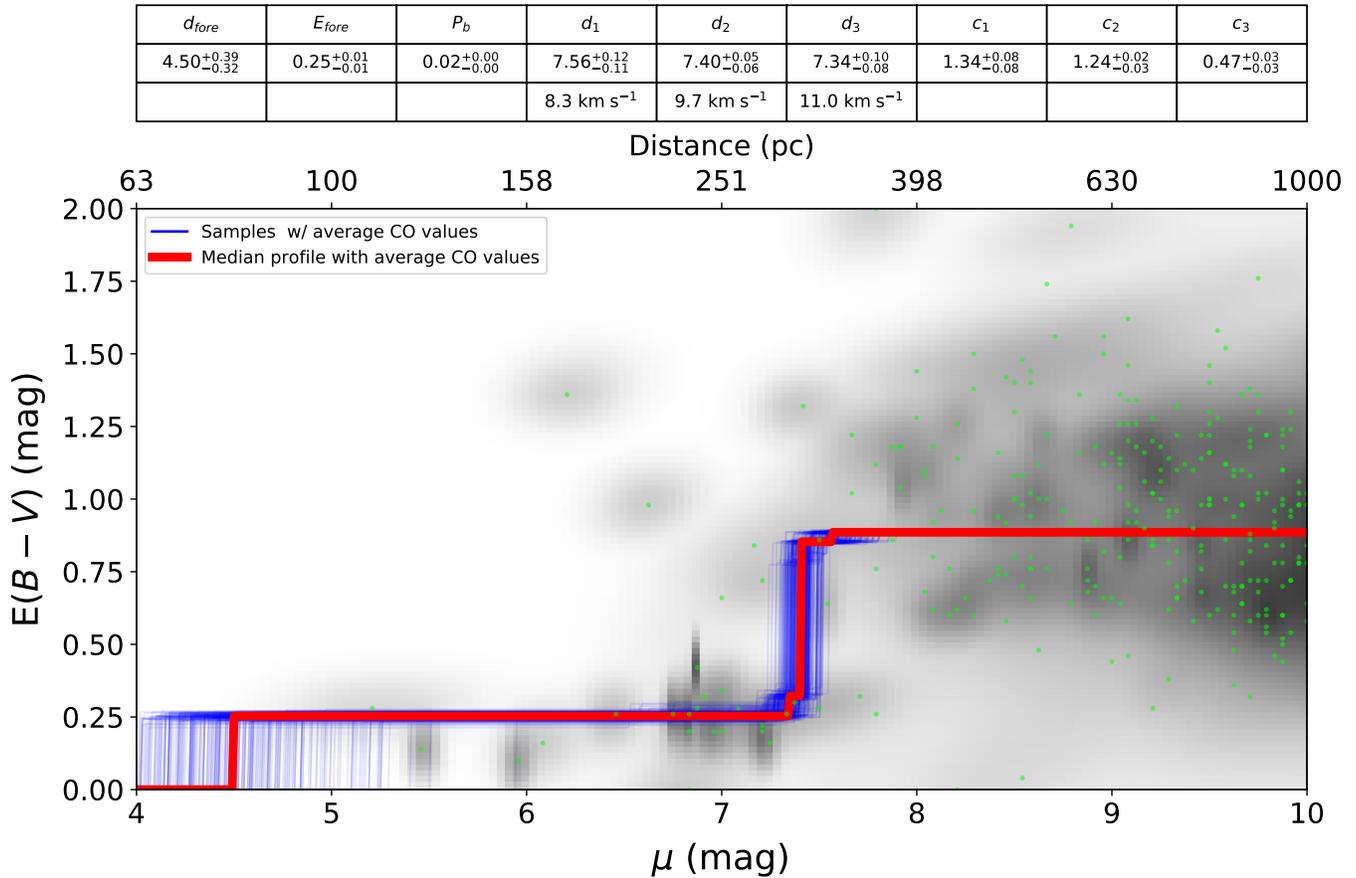

\begin{center}
\includegraphics[width=1.0\columnwidth]{{{B5_outlier_FINALPUB_profile}}}
\caption{{\label{fig:B5_profile} Line-of-sight reddening profile for the B5 star-forming region. The full profile is integrated out to $\mu=19$ mag (see Figures \ref{fig:example_profile} and \ref{fig:video}), but for illustrative purposes we only show the reddening profile in the range $\rm 4 \; mag \; < \mu < 10 \; mag$. The red line indicates the reddening profile parameterized by the median of the samples for each parameter (summarized in the table above the figure), derived from our \texttt{dynesty} run. Since the magnitude of the reddening jump depends on the amount of CO emission, we have taken the average of the CO emission in each slice to obtain an estimate of the reddening to B5. The blue lines are random samples from the same chain used to derive the median. In the background grayscale, we show distance-reddening stellar posteriors used in the parameter estimation problem (see \citetalias{Green_2018}), stacked on top of one another. We plot the most probable distance-reddening position for each star as a lime green scatter point, obtained by extracting the cell with the maximum probability in each of the gridded stellar posteriors. These lime green points are shown only for reference and are \textit{not} used in the fit, which always integrates through the full posterior for each star.}}
\end{center}
\end{figure}

The line-of-sight reddening profile for B5 determined using the median of the samples from our \texttt{dynesty} run is shown in red in Figure \ref{fig:B5_profile}. The line-of-sight reddening profiles determined by drawing random weighted samples from the same run are shown in blue, which gives a sense of the underlying uncertainty in the parameters. While the CO intensity (and thus the magnitude of the reddening jump corresponding to each slice) changes from pixel to pixel, for illustrative purposes we take the average CO value of each velocity slice to draw the profiles. In the background grayscale, we show the stacked stellar posteriors of distance and reddening for all the stars used in the analysis. Finally, we overlay the most probable distance and reddening for each star in lime green, obtained by extracting the cell in each gridded stellar posterior array with the maximum probability. 

We place B5 about 50 pc further away than \citet{Cernis_1993}. That study performs optical photometry on dozens of stars in the vicinity of IC348 (including B5). From this multi-band optical photometry \citet{Cernis_1993} infers a spectral type and intrinsic color for each star by assuming some extinction curve and determining where the stars' unreddened colors intersect the stellar locus of main sequence stars in various color-color projections.  The reddening then follows from the difference between the star's observed and intrinsic colors, which can subsequently be used to determine the distance to the stars through the adoption of a fixed $R(V)$ value. \citet{Cernis_1993} is able to roughly bracket B5 between lower extinction foreground stars and higher extinction background stars, constraining the distance to B5 to $\approx \rm 250-270 \; pc$, which is significantly lower than our distances of $d \approx 300$ pc. However, considering that \citet{Cernis_1993} states their uncertainties can be as high as 25\%, our two distances can still be reconciled. 

Our distances to B5 agree well with the distances to the same region from \citet{Schlafly_2014}, which are derived using a technique similar to the one presented in this work. Recall that \citet{Schlafly_2014} finds distances to $0.2^\circ$ sightlines distributed systematically across the Perseus Molecular Cloud, by modeling the line-of-sight reddening distribution as caused by a single dust cloud at some distance $d$, with an angular distribution given by Planck \citep{Planck_2011}. \citet{Schlafly_2014} then samples the most probable distance to the cloud, by determining which reddening profile is most consistent with distance and reddening posteriors \citep{Green_2014,Green_2015} for stars along the same line of sight. Considering sightlines in the immediate vicinity of B5 ($< 0.1 ^\circ$ from our region of interest shown in Figure \ref{fig:perseus_regions}), with uncertainties $<$ 20\%, \citet{Schlafly_2014} determines distances of $278^{+34}_{-25}$ pc, $321^{+24}_{-24}$ pc, and $352^{+53}_{-50}$ pc so they also favor higher distances on average than \citet{Cernis_1993}. 

\subsection{IC348}
We consider four velocity slices towards IC348 (with CO intensity-weighted velocities of 5.4, 7.4, 9.0, and $10.4 \; \rm km \; s^{-1}$) and determine $\mu$ of $7.35^{+0.05}_{-0.04}$, $7.35^{+0.04}_{-0.03}$, $7.34^{+0.04}_{-0.03}$, and $7.35^{+0.05}_{-0.03}$ mag, respectively. This corresponds to distances of $295^{+7}_{-5}$, $295^{+5}_{-4}$, $294^{+5}_{-4}$, and $295^{+7}_{-4}$ pc. The different velocity components of IC348 correspond to nearly identical distances, and all the slices are consistent with being at the same distance given the uncertainty on each slice. We find an average distance to IC348 of $\mu = 7.35 \pm 0.03$ mag ($295 \pm 4$ pc). The line-of-sight reddening profile for IC348 determined using the median of the samples from our \texttt{dynesty} run is shown in red in Figure \ref{fig:IC348_profile}. 

The distance to IC348 has been much debated in the literature since the 1950's. Almost all distances are photometric and rely on directly or indirectly determining the color excess in $B-V$, which can then be used to calculate the distance similarly to the \citet{Cernis_1990, Cernis_1993} method. Like B5, \citet{Cernis_1990} places IC348 at a distance of $\approx \rm 250-270 \; pc$. However, similar photometric-based studies by \citet{Trullols_1997} and \citet{Harris_1954} determine distances to IC348 of 240 pc and 316 pc, respectively, so there is little agreement in the literature. Our distances to IC348 are in agreement with the distance of $297^{+43}_{-28}$ pc that \citet{Schlafly_2014} calculates for a sightline intersecting IC348 (again, derived using a very similar technique as presented in this work). The most recent constraint comes from \citet{Ortiz_Leon_2018}, which determines a distance to the cloud using both Gaia DR2 parallax measurements and VLBA astrometric observations of young stars associated with the IC348 cluster. \citet{Ortiz_Leon_2018} finds a distance of $320 \pm 26$ pc using the Gaia DR2 data and $321 \pm 10$ pc using the VLBA observations. While \citet{Ortiz_Leon_2018} places the cloud at a slightly farther distance than this work, our results are consistent given the quoted uncertainties. 

\begin{figure}[h!]
\begin{center}
\includegraphics[width=1.0\columnwidth]{{{IC348_outlier_FINALPUB_profile}}}
\caption{{\label{fig:IC348_profile} Line-of-sight reddening profile for the IC348 star-forming region in the range $4 < \mu < 10$ mag. The meaning of points, lines, and the background grayscale are the same as in Figure \ref{fig:B5_profile}. }}
\end{center}
\end{figure}

\subsection{B1}
We consider five velocity slices towards B1 (with CO intensity-weighted velocities of 1.1, 2.8, 4.8, 6.7, and 8.2 $\rm km \; s^{-1}$) and determine $\mu$ of $7.13^{+0.22}_{-0.25}$,  $7.36^{+0.16}_{-0.23}$, $7.48^{+0.04}_{-0.03}$, $7.38^{+0.03}_{-0.03}$, and $7.45^{+0.04}_{-0.06}$ mag, respectively. This corresponds to distances of $266^{+28}_{-32}$, $297^{+23}_{-33}$, $313^{+6}_{-5}$, $300^{+4}_{-4}$, and $309^{+5}_{-8}$ pc. The velocity slices towards B1 show little dispersion in distance, with four out of five slices within $\mu \lesssim 0.1$ mag of each other. We find an average distance to B1 of $\mu = 7.39 \pm 0.03$ ($301 \pm 4$ pc). The line-of-sight reddening profile for B1 determined using the median of the samples from our \texttt{dynesty} run is shown in red in Figure \ref{fig:B1_profile}.

Again, using the same technique as implemented in \citet{Cernis_1993} for B5 and IC348, \citet{Cernis_2003} determines a distance of $d=230$ pc to B1 by calculating the extinction and distance to 98 stars towards the B1 sightline. Like B5 and IC348, we find that our distances to B1 are systematically further away, this time on the order of 70 pc.  

\begin{figure}[h!]
\begin{center}
\includegraphics[width=1.0\columnwidth]{{{B1_outlier_FINALPUB_profile}}}
\caption{{\label{fig:B1_profile} Line-of-sight reddening profile for the B1 star-forming region in the range $4 < \mu < 10$. The meaning of points, lines, and the background grayscale are the same as in Figure \ref{fig:B5_profile}. }}
\end{center}
\end{figure}

\subsection{NGC1333}
\begin{figure}[h!]
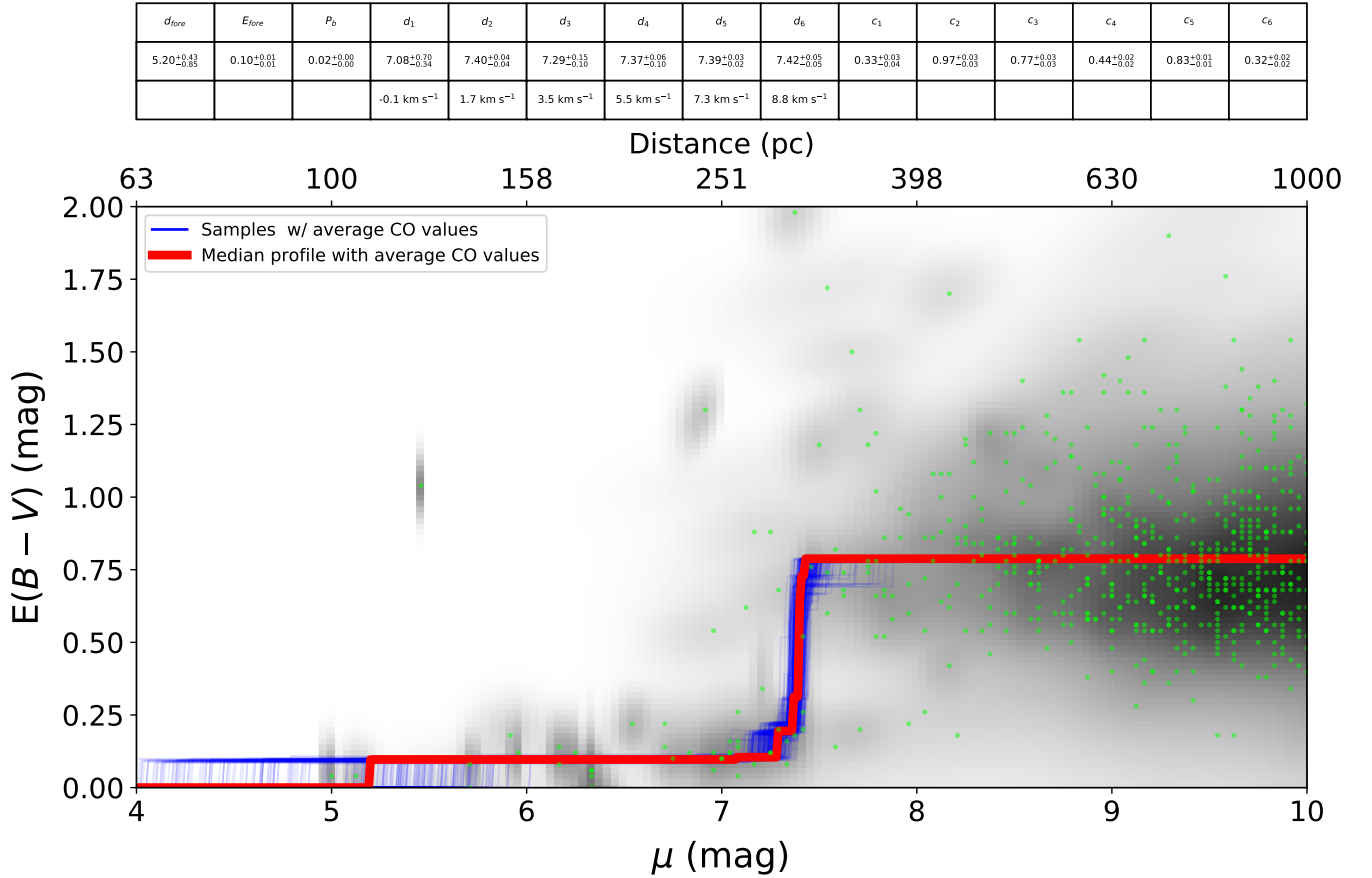

\begin{center}
\includegraphics[width=1.0\columnwidth]{{{NGC1333_outlier_FINALPUB_profile}}}
\caption{{\label{fig:NGC1333_profile} Line-of-sight reddening profile for the NGC1333 star-forming region in the range $4 < \mu < 10$ mag. The meaning of points, lines, and the background grayscale are the same as in Figure \ref{fig:B5_profile}.}}
\end{center}
\end{figure}

We consider six velocity slices towards NGC1333 (with CO intensity-weighted velocities of -0.1, 1.7, 3.5, 5.5, 7.3, and 8.8 $\; \rm km \; s^{-1}$) and determine $\mu$ of $7.08^{+0.70}_{-0.34}$, $7.40^{+0.04}_{-0.04}$, $7.29^{+0.15}_{-0.10}$,  $7.37^{+0.06}_{-0.10}$, $7.39^{+0.03}_{-0.02}$ and $7.42^{+0.05}_{-0.05}$ mag, respectively. This corresponds to distances of $260^{+99}_{-44}$, $302^{+5}_{-6}$, $287^{+20}_{-13}$, $298^{+9}_{-14}$, $301^{+4}_{-3}$, and $305^{+7}_{-7}$ pc. We find an average distance to NGC1333 of $\mu = 7.38 \pm 0.02$ ($299 \pm 3$ pc). While most of the slices are well-constrained and at a similar distance ($\approx$ 280-300 pc), we find that the distance to the first slice, at the lowest velocity, lies at a distance of around 260 pc.  Looking at a 3D volume rendering of the CO emission in NGC1333 (from e.g. the \href{https://www.cfa.harvard.edu/COMPLETE/movies.html}{CO COMPLETE survey page}) we see that there is actually a lower velocity wisp-like structure in front of the bulk of the higher velocity CO emission constituting this cloud. We find that the velocity of our first slice, $v_1$, is spatially and kinematically coincident with this wispy structure in front of NGC1333, suggesting our algorithm is able to place (albeit with larger uncertainties) this structure at a different distance than the bulk of the cloud. The larger uncertainties are due to the fact that these slices are contributing only a small amount of CO emission to the total along the line of sight. Thus, they make weaker dust templates, leading to more poorly constrained distances. Nevertheless, it is promising that our method is able to separate this wisp structure from the rest of the cloud. The line-of-sight reddening profile for NGC1333 determined using the median of the samples from our \texttt{dynesty} run is shown in red in Figure \ref{fig:NGC1333_profile}. 

In comparison to the literature, we place NGC1333 on average 60 pc further away than \citet{Hirota_2008}, which determines a distance of $\rm 235 \pm 18 \; pc$ to NGC1333 using astrometry of $\rm H_2O$ masers associated with a young stellar object (YSO) embedded inside the cloud. We note that the \citet{Hirota_2008} result is in agreement with the photometric distance to NGC1333 from \citet{Cernis_1990} (220 pc), which uses the same technique as implemented in \citet{Cernis_1993} for B5. \citet{Hirota_2008} estimates their uncertainties to be on the order of 8\%, in comparison to the 25\% uncertainty quoted by \citet{Cernis_1990}. We find the typical uncertainty on our average distance to be about 1-2\%. However, we reiterate that this only accounts for statistical uncertainty and neglects any systematic uncertainty, which we estimate to be $\mu=0.1$ mag or $5\%$ in distance (see \S \ref{results}). When added in quadrature, this produces total combined uncertainties on the order of $\approx 5-6 \%$. The Gaia DR2 astrometric data release \citep{Lindegren_2018} supports a farther distance to the Perseus complex as a whole, which we discuss further in \S \ref{discussion}. As in the case of B5 and IC348, the \citet{Schlafly_2014} distance to a sightline near NGC1333 (d=$288^{+39}_{-29}$ pc) is in agreement with our CO-based distances. Finally, more recently, \citet{Ortiz_Leon_2018} finds a distance to the cloud using Gaia DR2 parallax measurements of young stars associated with the NGC1333 cluster, obtaining a value of $293 \pm 22$ pc. This is in strong agreement with the average distance to NGC1333 we present in this work. 

\subsection{L1448}
\begin{figure}[h!]
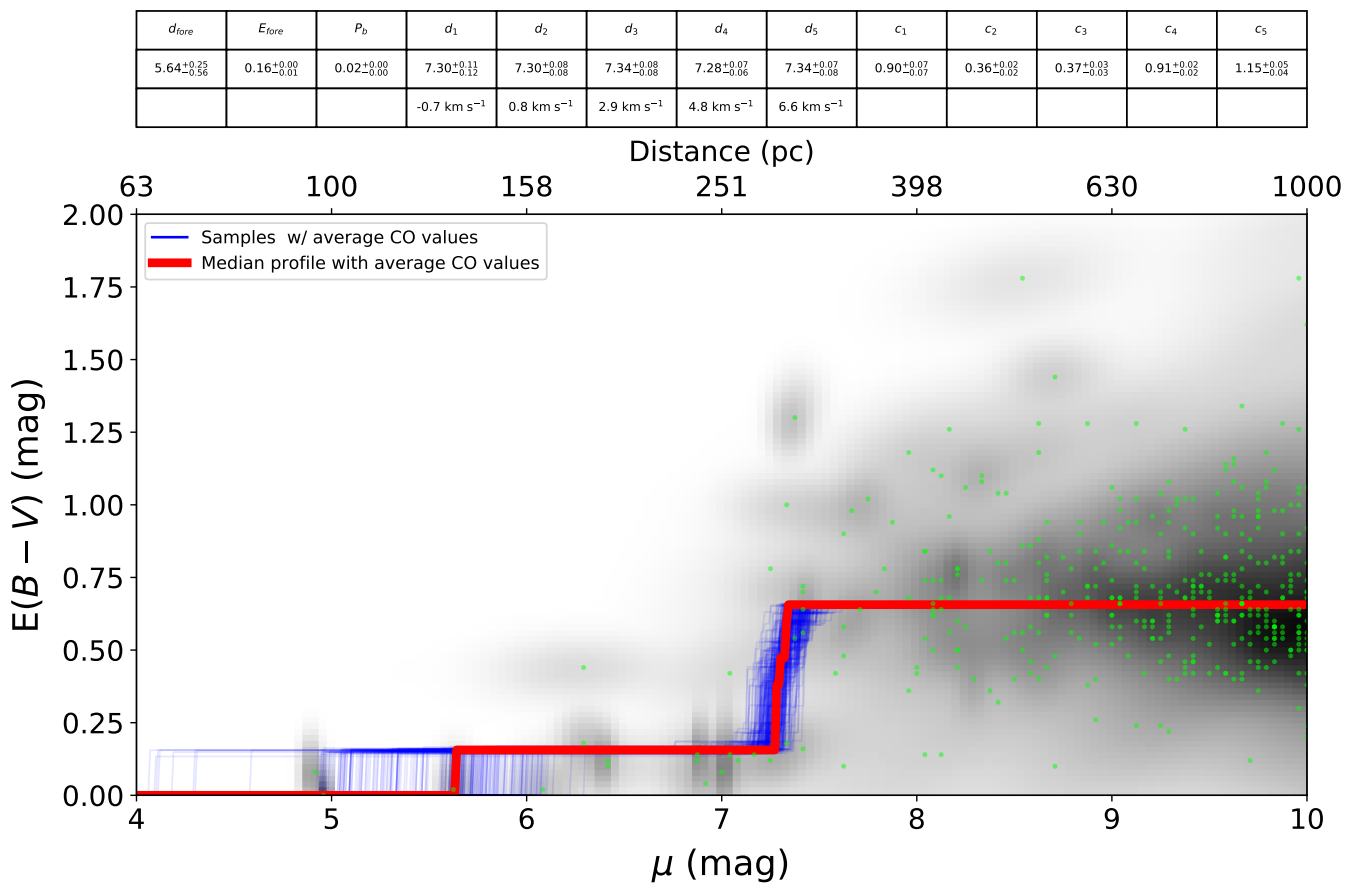

\begin{center}
\includegraphics[width=1.0\columnwidth]{{{L1448_outlier_FINALPUB_profile}}}
\caption{{\label{fig:L1448_profile} Line-of-sight reddening profile for the L1448 star-forming region in the range $4 < \mu < 10$ mag. The meaning of points, lines, and the background grayscale are the same as in Figure \ref{fig:B5_profile}. }}
\end{center}
\end{figure}

We consider five velocity slices towards L1448 (with CO intensity-weighted velocities of -0.7, 0.8, 2.9, 4.8, and 6.6 $\rm km \; s^{-1}$) and determine $\mu$ of $7.30^{+0.11}_{-0.12}$, $7.30^{+0.08}_{-0.08}$, $7.34^{+0.08}_{-0.08}$, $7.28^{+0.07}_{-0.06}$, and $7.34^{+0.07}_{-0.08}$ mag, respectively. This corresponds to distances of $288^{+15}_{-17}$, $289^{+10}_{-11}$, $294^{+11}_{-12}$, $286^{+9}_{-8}$, and $293^{+10}_{-11}$ pc. All of the velocity slices are consistent with being at the average distance to the cloud, which we find to be $\mu = 7.30 \pm 0.04$ ($288 \pm 6$ pc). The line-of-sight reddening profile for L1448 determined using the median of the samples from our \texttt{dynesty} run is shown in red in Figure \ref{fig:L1448_profile}.

As is also the case with NGC1333, we place L1448 approximately 50 pc further away than the parallax distance derived from maser parallax measurements. Using a similar technique to \citet{Hirota_2008}, \citet{Hirota_2011} monitor maser activity associated with a YSO embedded in the L1448 star-forming region, and determine a parallax distance of $\rm 232 \pm 18 \; pc$.  Finally, we note that \citet{Schlafly_2014} also prefer a farther distance to L1448, placing the cloud at a distance of $261^{+36}_{-43}$ pc, which is $<$ 20 pc from the average distance we find for the cloud. 

\subsection{L1451}
\begin{figure}[h!]
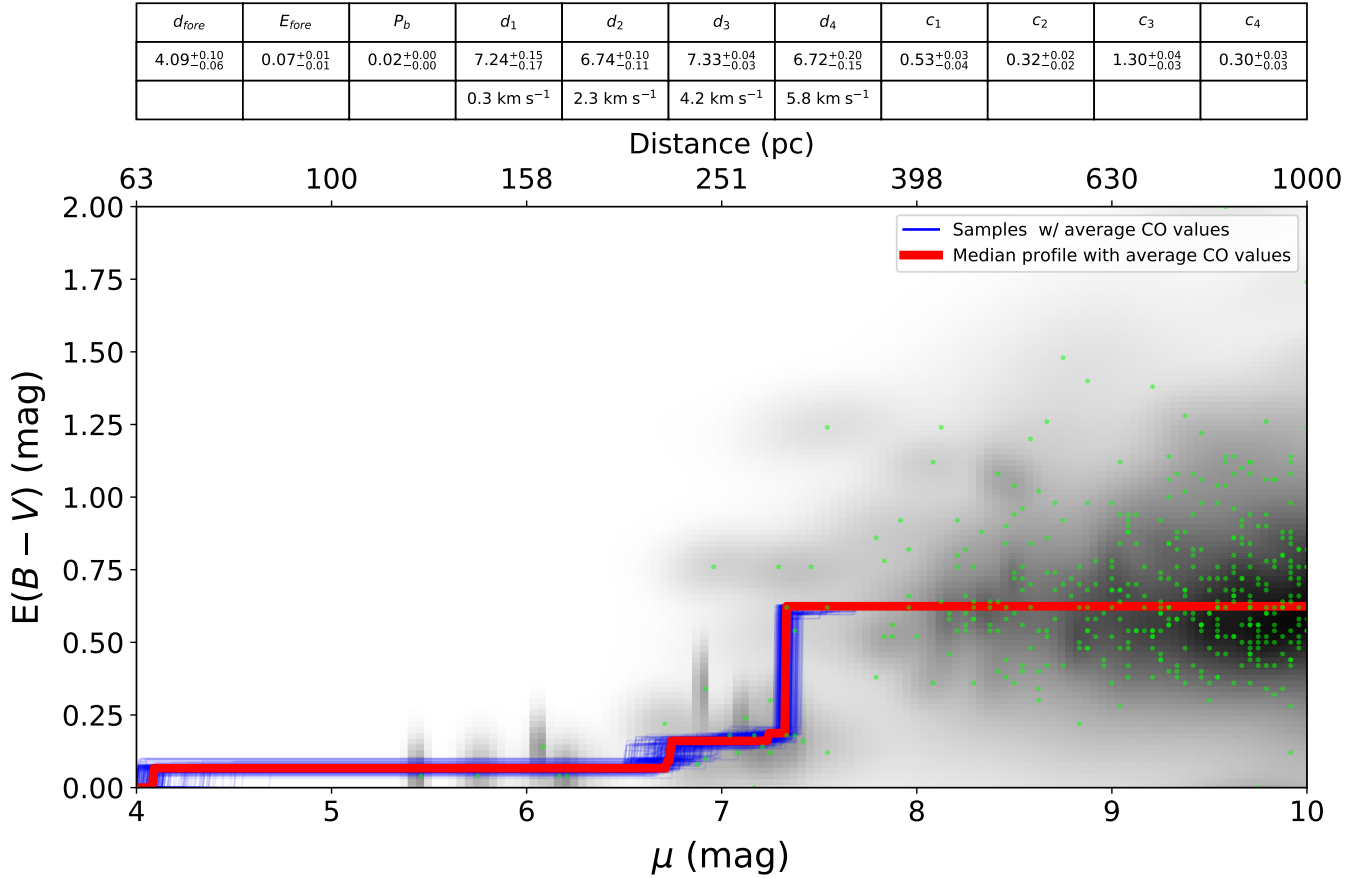

\begin{center}
\includegraphics[width=1.0\columnwidth]{{{L1451_outlier_FINALPUB_profile}}}
\caption{{\label{fig:L1451_profile} Line-of-sight reddening profile for the L1451 star-forming region in the range $4 < \mu < 10$ mag. The meaning of points, lines, and the background grayscale are the same as in Figure \ref{fig:B5_profile}.}}
\end{center}
\end{figure}

We consider four velocity slices towards L1451 (with CO intensity-weighted velocities of 0.3, 2.3, 4.2, and 5.8 $\rm km \; s^{-1}$) and determine $\mu$ of $7.24^{+0.15}_{-0.17}$, $6.74^{+0.10}_{-0.11}$,  $7.33^{+0.04}_{-0.03}$, and $6.72^{+0.20}_{-0.15}$ mag, respectively. This corresponds to distances of $280^{+20}_{-22}$, $223^{+11}_{-11}$, $292^{+6}_{-4}$, and $220^{+21}_{-16}$ pc, respectively. The velocity slice corresponding to the highest amount of reddening lies at 292 pc, but a non-negligible amount of its dust also lies at closer distance---between about 220 and 280 pc. We find an average distance to L1451 of $\mu = 7.23 \pm 0.03$ ($279 \pm 4$ pc). The average distance to L1451 agrees well with the average distance to L1448 ($288 \pm 6$ pc), which lie in close proximity spatially on the plane of the sky. The line-of-sight reddening profile for L1451 determined using the median of the samples from our \texttt{dynesty} run is shown in red in Figure \ref{fig:L1451_profile}.

In comparison to the other regions discussed, there have been fewer attempts to determine a distance to L1451. Most studies which target L1451 either assume an average distance to the cloud based on a compilation of previous distance estimates for the entire Perseus complex \citep[$\approx$ 250 pc,][]{Pineda_2011} or else assign the same distance determined via trigonometric parallax measurements for the neighboring L1448 and NGC1333 regions  \citep[$\approx$ 230 pc,][]{Storm_2016, Maureira_2017}. The most targeted distance estimate to the cloud comes from \citet{Schlafly_2014}, which considers a sightline about half a degree away from the L1451-mm dense core \citep[see][]{Pineda_2008}. For that sightline, \citet{Schlafly_2014} determines a distance of $266^{+27}_{-31}$ pc, which is in good agreement with our average distance. 

\section{Discussion} \label{discussion}
Across the entire Perseus Complex, the distances to our velocity slices typically vary between $d\approx 260-310$ pc. The slices constituting the bulk of the reddening for each cloud typically lie within $\mu \lesssim 0.1$ mag of each other, while the overall dispersion across the entire cloud is on the order of $\mu \approx 0.2$ mag. From east to west we find the average distance to B5, IC348, B1, NGC1333, L1448, and L1451 to be 302 pc, 295 pc, 301 pc, 299 pc, 288 pc, and 279 pc, respectively. We summarize our results in Figure \ref{fig:perseus_master}, which includes plots showing how the Declination (panel a), velocity (panel b) and average reddening-weighted distance (panel c) for each region varies as a function of Right Ascension, as well as how the cloud  Declination (panel d) and average reddening-weighted distance (panel e) varies as a function of cloud velocity. The background colorscale in panels (a), (c), and (d) are different projections of a combined volume rendering of the $\rm ^{12}CO$ (red) and $\rm ^{13}CO$ (blue) Perseus COMPLETE cubes. A movie showing the full 3D volume rendering of this cube can be found on the \href{https://dataverse.harvard.edu/dataset.xhtml?persistentId=doi%3A10.7910%2FDVN%2FLWYIJ3}{Dataverse}.\footnote{doi:10.7910/DVN/LWYIJ3}

In more detail, we discuss how velocity correlates with distance in Perseus in \S \ref{velocities_to_distances} and how our CO-reddening-based distance estimates compare to those derived from maser parallax observations in \S \ref{parallax_discussion}. 

\subsection{Mapping Velocities to Distances} \label{velocities_to_distances}
\begin{figure}[h!]
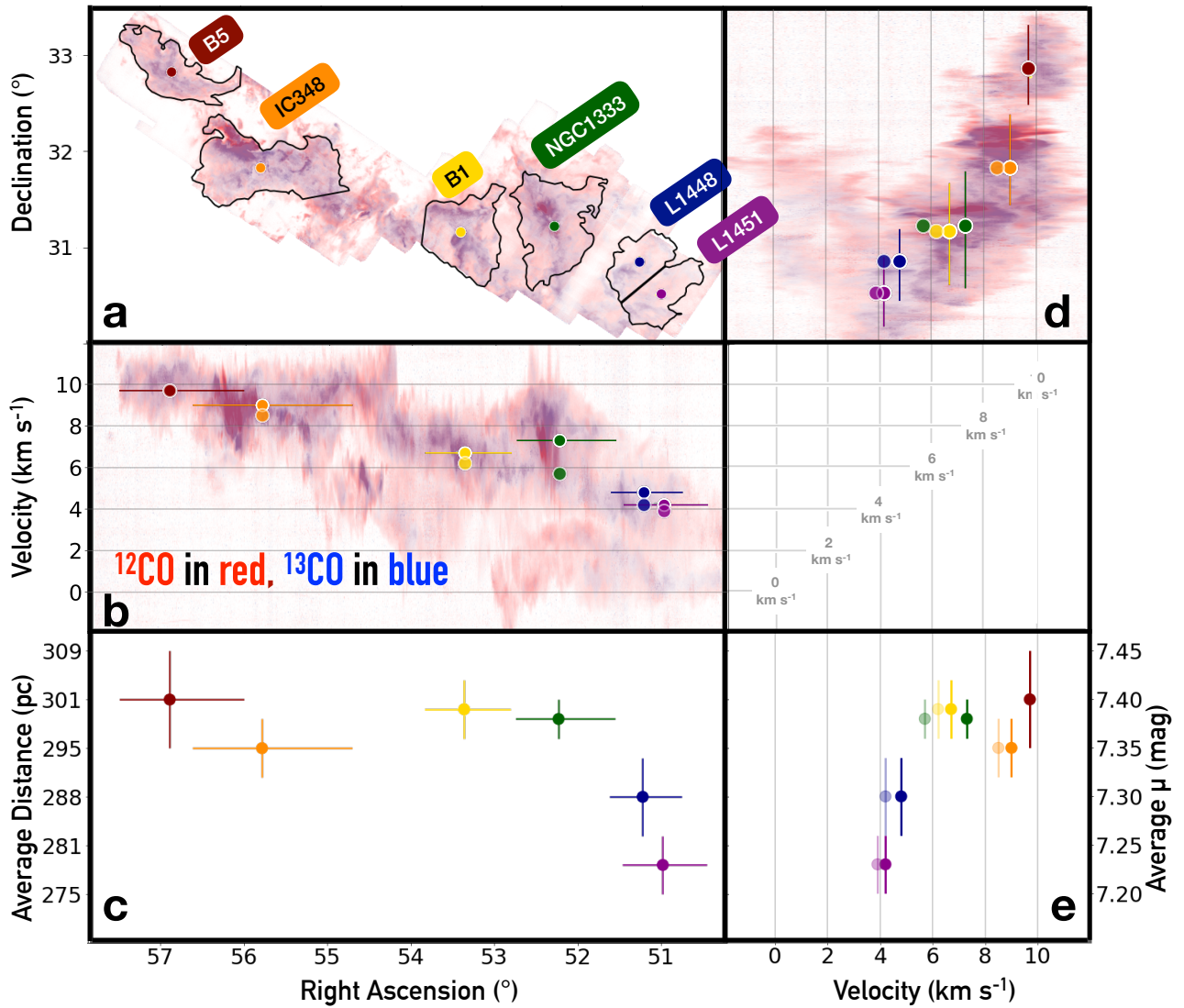

\begin{center}
\includegraphics[width=1.0\columnwidth]{{{Perseus_Master_FINAL}}}
\caption{{\label{fig:perseus_master} \textbf{Panel a}: Combined $\rm ^{12}CO$ (red) and $\rm ^{13}CO$ (blue) integrated intensity map of Perseus \citep{Ridge_2006}. The boundaries we define for each region (same as in Figure \ref{fig:perseus_regions}b) are shown in black. The centroid of each polygon is marked with a different colored point. \textbf{Panel b}: an \textit{RA-velocity} diagram of Perseus, with the same colorscale as panel (a). The colored points show the peak-reddening velocity (dark points) and average-reddening weighted velocity (light points) as a function of Right Ascension for each cloud. The errorbars in Right Ascension show the horizontal extents of the polygons overlaid in panel (a). \textbf{Panel c}: Average reddening-weighted distance to each region as a function of Right Ascension. \textbf{Panel d}:  a \textit{velocity-DEC} diagram of Perseus, with the same colorscale as panel (a). The colored points show the Declination as a function of peak reddening velocity (dark points) and average-reddening weighted velocity (light points). The errorbars in Declination show the vertical extents of the polygons overlaid in panel (a). \textbf{Panel e:} The average reddening-weighted distance to each region as a function of its peak reddening velocity (dark points) and average reddening-weighted velocity (light points). We find that the cloud distances tend to increase with both average reddening-weighted velocity and Right Ascension. In total, the velocity gradient of $\rm \approx 5 \; km \; s^{-1}$ maps to a distance gradient of about 25 pc. The uncertainty provided on distance only accounts for the statistical uncertainty and does not include any systematic uncertainty, which we estimate to be 0.1 mag in distance modulus or 5\% in distance (see \S \ref{results}).}}
\end{center}
\end{figure}

Because we sample for the distances to the velocity slices of CO spectral cubes, we can \textit{explicitly tie our distance measurements to the velocity structure of the molecular gas}. Thus, for the first time, we can show that the velocity gradient from east to west roughly maps to a corresponding distance gradient. This distance gradient is apparent in Figure \ref{fig:perseus_master}e, which shows average distance as a function of both peak-reddening velocity (dark colored points) and average reddening-weighted velocity (light colored points). We note that a potential distance gradient has been proposed before \citep[for instance, in][]{Ridge_2006} but could not be confirmed with any confidence due to the fact that our traditional method of translating molecular clouds' line-of-sight velocities to distances \citep[via a Galactic rotation curve e.g.][]{Roman_Duval_2009} cannot be applied locally because peculiar motions dominate over the motions due to Galactic rotation at such close separation. 

In more detail, Figure \ref{fig:perseus_master}e shows that the average velocity of the slices towards B5 clouds lies at $\rm 10 \; km \; s^{-1}$, while the slices towards L1451 lie at $\rm 4 \; km \; s^{-1}$. As a result, the typical velocity gradient of $\rm \approx 5 \; km \; s^{-1}$ across the cloud translates to a distance gradient of 25 pc. There is a clear trend --- cloud with higher average velocities tend to have higher average distances. 

To put this distance gradient in context, the angular length of Perseus spans $\approx 5^\circ$ on the plane of the sky, which corresponds to a project length of $\approx 25$ pc assuming our average distance to the complex (294 pc). This means that the line-of-sight extent of Perseus is likely equal to its projected length, giving it an aspect ratio between $1:1$ and $2:1$. This also suggests Perseus is inclined to our line of sight by $\approx 45^\circ$. Given that most profiles do not show a large dispersion in distance along the line-of-sight (particularly for those slices which contain high amounts of reddening), it can be reasonably assumed that the cloud is relatively compact in depth, and that it does not have a ``sheet"-like morphology, as has been proposed for clouds like Musca \citep{Tassis_2018}. 

While a single cloud distance for all of Perseus---usually around 250 pc---is often adopted \citep[e.g.][]{Curtis_2009, Arce_2010, Enoch_2006, Rebull_2007, Campbell_2016} our results suggest that this may be inappropriate. Again, this potential distance gradient has been discussed extensively in the literature (including in many of the cloud-wide studies that adopt a single cloud distance); however, the distance gradient has been difficult to implement due to the diversity of distance mapping techniques and the high uncertainties associated with certain methods \citep[e.g. the photometric distance method from][]{Cernis_1990,Cernis_1993, Cernis_2003}. Nevertheless, since properties like clump mass are dependent on the distance squared, distance discrepancies of just a few tens of parsecs could produce variations in these properties as high as 50\%. Our CO-reddening based distances are systematically calculated across the cloud and have typical combined statistical, sampling, and systematic uncertainties $\approx 5\%$, so they can be used in lieu of traditional photometric methods to constrain the distances to star-forming regions in the Perseus Molecular Cloud that lie at different velocities. 

\subsection{Perseus is Farther than we Thought} \label{parallax_discussion}
\begin{figure}[h!]
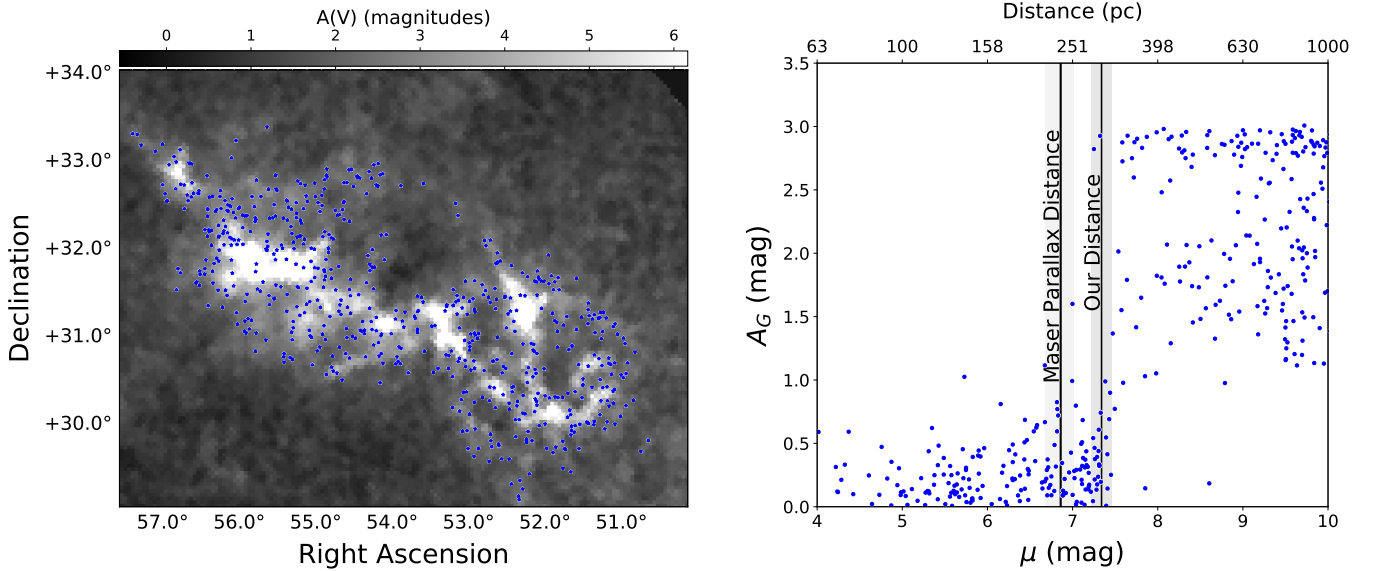

\begin{center}
\includegraphics[width=1.0\columnwidth]{{{Gaia_Agreement_Revision}}}
\caption{{\label{fig:gaia_validation} Validating our results through a comparison to Gaia G-band extinction ($A_G$) and parallax measurements for a reliable sample of Gaia DR2 stars along reddened sight lines throughout Perseus (see \S \ref{parallax_discussion} for more details on the quality cuts we make). In the left panel, we show the sample overlaid in blue on a NICER extinction map of Perseus \citep[same as shown in Figure \ref{fig:perseus_regions};][]{Pineda_2008}. On the right we show a plot of $A_G$ versus parallax-based distance for these same stars. A visible step in reddening, attributable to Perseus, can be seen. Also in the right panel, we show that our average distance to the cloud (294 pc) agrees better with Gaia, and that previous maser-parallax based distances to the cloud \citep[230 pc;][]{Hirota_2008, Hirota_2011} likely underestimate its distance. The uncertainties on each distance are shown via the gray shaded regions.}}
\end{center}
\end{figure}

Our distances to the NGC1333 and L1448 star-forming regions are about 60 pc farther away than those derived via trigonometric parallax measurements of water masers from \citet{Hirota_2008, Hirota_2011}. \citet{Hirota_2008, Hirota_2011} determine a maser-based distance of $d\approx 230$ pc to the NGC1333 and L1448 star-forming regions, and they estimate their uncertainties to be $\approx 8\%$. This is the typical distance for Perseus adopted in the literature \citep[see e.g.][]{Lee_2015}. The distances we determine for L1448 and NGC1333 (288 and 299 pc, respectively) are consistent with the higher overall average distance we compute for the cloud as a whole (294 pc, see Table \ref{average_distances}). Gaia DR2 parallax and G-band extinction ($A_G$) measurements towards stars throughout Perseus strongly favor the higher distances to the complex that we present in this work, and act as further validation for this method. This is illustrated in Figure \ref{fig:gaia_validation}. For this comparison, we select a subsample of Gaia DR2 stars along reddened sight lines throughout Perseus and we require that they have reliable measurements for both parallax and $A_G$. Specifically, we impose the criteria that that the stars must lie in pixels with total line-of-sight visual extinction $A(V) > 2$ mag (from the NICER map, see also the left panel of Figure \ref{fig:perseus_regions}). Of these stars, we require their \texttt{parallax\_over\_error} to be greater than five, and that their $\pm 1\sigma$ uncertainties on $A_G$ (\texttt{a\_g\_percentile\_lower} and \texttt{a\_g\_percentile\_upper}) be within 0.25 mag of the median $A_G$ value (\texttt{a\_g\_val}) reported. These stars are shown overlaid in blue on the NICER extinction map of Perseus in the left panel of Figure \ref{fig:gaia_validation}; a plot of $A_G$ versus parallax-based distance for the same stars is shown in the right panel of Figure  \ref{fig:gaia_validation}. Also in the right panel, we overlay the average distance we compute for Perseus (294 pc, see Table \ref{average_distances}), along with the $\approx 230$ pc distance often adopted to the cloud based on the results from \citet{Hirota_2008, Hirota_2011}. The farther distance is clearly favored by the sample of Gaia stars most likely to accurately trace the jump in reddening attributable to the Perseus complex. If anything, we could be slightly underestimating the average distance to the Complex, as the Gaia data indicate that the average jump in reddening could lie anywhere from our reported average distance, up to $\approx 1-2$ standard deviations beyond it. 

\section{Conclusion} \label{conclusion}
We present a catalog of distances to major star-forming regions in the Perseus Molecular Cloud in the velocity range $\approx$ $-1$ to $\rm 12 \; km \; s^{-1}$. We produce the catalog using a two-step process. First, we infer the distance-reddening posteriors for batches of stars across Perseus based on the technique presented in \citetalias{Green_2018}, and we apply a Gaia parallax-based Gaussian likelihood term as an additional distance constraint. Then, we model the reddening along the line of sight towards these stars as a linear combination of the optical depth of CO velocity slices. The result is a set of distances tied to the velocity slices defining the structure of the molecular gas towards these clouds, which we then sample using a Monte Carlo method. We target the B5, IC348, B1, NGC1333, L1448, and L1451 star-forming regions, and find typical cloud distances of 302 pc, 295 pc, 301 pc, 299 pc, 288 pc, and 279 pc, respectively. On average, the $\rm 5 \; km \; s^{-1}$ velocity gradient maps to a corresponding distance gradient of $\approx 25$ pc, with the eastern half of Perseus systematically farther away than the western half. We find an average distance to the Perseus Complex as a whole of $294 \pm 17$ pc.

Typical uncertainties on our distances are on the order of $5\%$, with $1-2\%$ of that due to statistical uncertainties, and 5\% due to systematics.\footnote{We add the statistical and systematic uncertainties in quadrature to estimate the combined $\approx$ 5\% uncertainty on our distances} We place the western portion of the Perseus Molecular Cloud (NGC1333 and L1448) approximately 60 pc ($\approx 1 \sigma$) further away than the distances derived from maser parallax measurements towards the same regions \citep{Hirota_2008, Hirota_2011}. When comparing to Gaia-based parallax and G-band extinction measurements towards highly-reddened lines of sight throughout Perseus, we find strong support for our higher distance. 

We have only scratched the surface of what is possible through the combination of stellar photometry, CO observations, and Bayesian statistics. The accuracy of our cloud distances are directly linked to the reliability of our per-star distance and reddening posteriors, which are derived from near-infrared and optical photometry (see \citetalias{Green_2018}). The addition of deeper near-infrared surveys \cite[e.g. UKIDSS;][]{Warren_2007}, the incorporation of more accurate stellar parallax measurements (from Gaia DR3 and beyond), and the advent of all-sky optical surveys like LSST will only better constrain stellar distances and reddenings, enabling us to probe deeper into dust-enshrouded regions like Perseus. Moreover, our technique is flexible enough to incorporate models with multiple clouds along the line of sight, which would prove useful towards the inner galaxy.  A similar technique \citep[known as ``kinetic tomography";][]{Tchernyshyov_2017, Tchernyshyov_2018} has already been applied across much of the sky, resulting in a four-dimensional (position-position-distance-velocity) reconstruction of the interstellar medium, albeit at lower distance resolution. Thus, the methodology presented here for the Perseus Molecular Cloud could be applied to local molecular clouds across the full sky, paving the way for accurate cloud distances (tied to the distribution of CO molecular gas) in the solar neighborhood and beyond.

\section{Acknowledgements}
We would like to thank Mark Reid for his expertise regarding the uncertainty surrounding distance measurements derived from maser parallax observations, Thomas Dame for his expertise on the structure and properties of CO molecular gas in the galaxy, and Charlie Conroy for insight regarding the stellar models we implemented to derive the per-star distance and reddening posteriors. All three also provided valuable comments on an early draft of this work. 

\pagebreak
\section{Appendix} 

\subsection{Dynesty Setup} \label{dynesty_appendix}
We use the following \texttt{dynesty} code to produce the chains used for parameter estimation in \S \ref{results}.
\begin{lstlisting}
    sampler = dynesty.NestedSampler(log-likelihood, prior-transform, ndim, bound=`multi', sample=`rwalk', update_interval=6., nlive=300, walks=50)
    sampler.run_nested(dlogz=0.1)
\end{lstlisting}

where ``log-likelihood" is our log-likelihood function (described in \S \ref{model}) and ``prior-transform" is a function that transforms our priors (described in \S \ref{model}) from an `ndim'-dimensional unit cube to the parameter space of interest. We set our convergence threshold, ``dlogz", equal to 0.1.

\begin{figure}[h!]
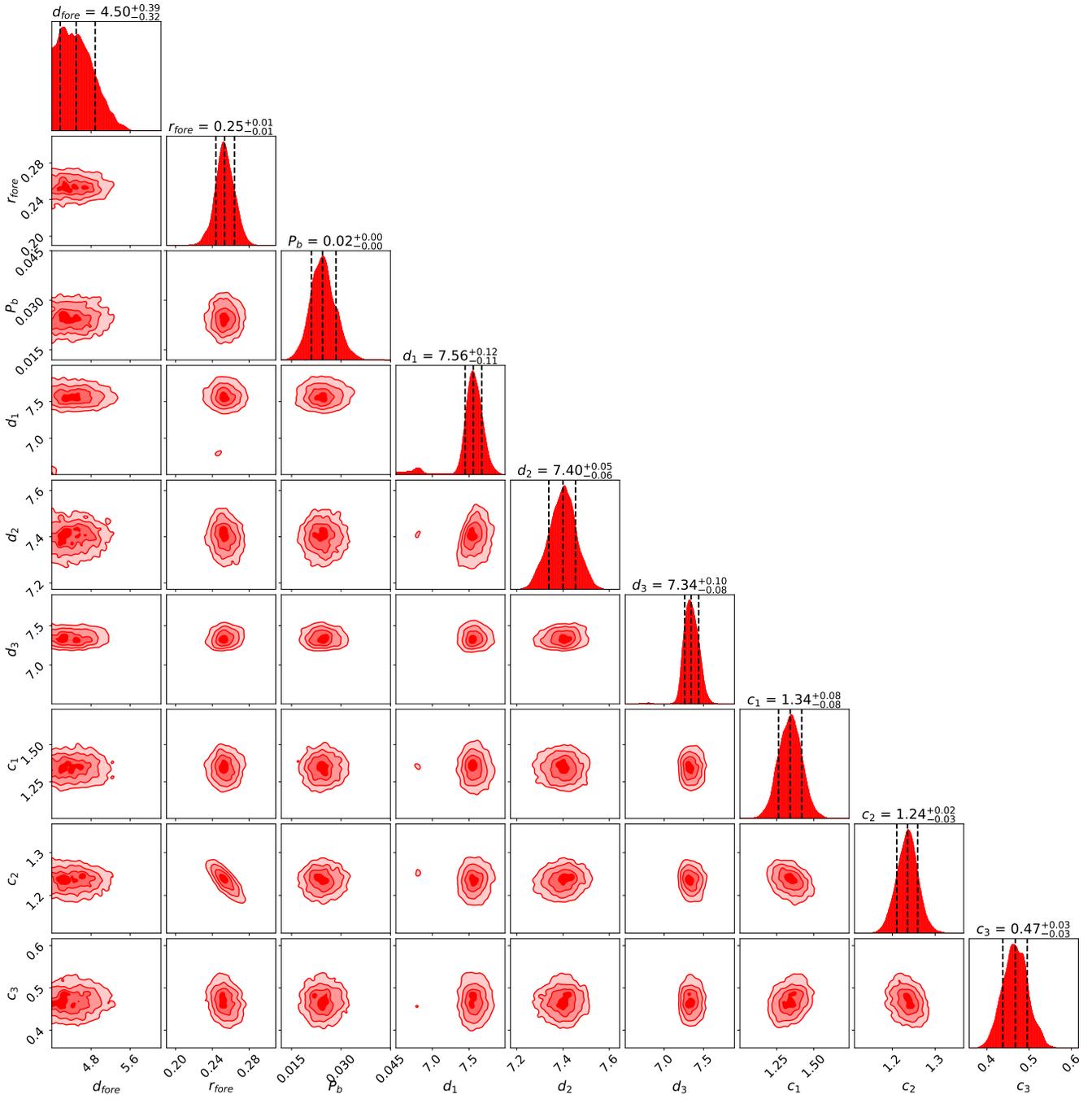

\begin{center}
\includegraphics[width=1.0\columnwidth]{{{B5_outlier_FINALPUB_corner}}}
\caption{{\label{fig:B5_cornerplot}  Corner plot derived from our \texttt{dynesty} run towards the B5 region.}}
\end{center}
\end{figure}

\begin{figure}[h!]
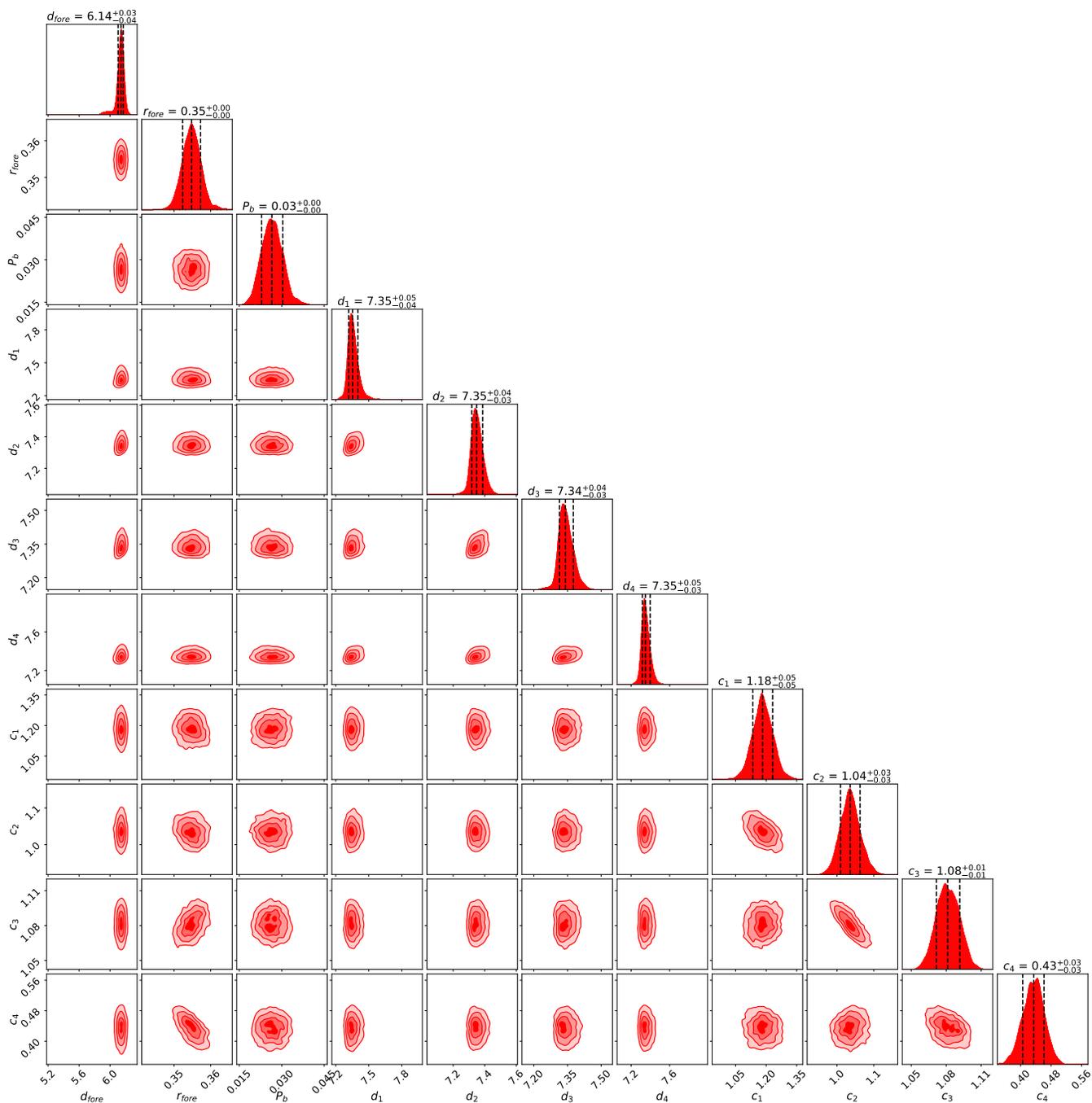

\begin{center}
\includegraphics[width=1.0\columnwidth]{{{IC348_outlier_FINALPUB_corner}}}
\caption{{\label{fig:IC348_cornerplot}  Corner plot derived from our \texttt{dynesty} run towards the IC348 region.}}
\end{center}
\end{figure}

\begin{figure}[h!]
\begin{center}
\includegraphics[width=1.0\columnwidth]{{{B1_outlier_FINALPUB_corner}}}
\caption{{\label{fig:B1_cornerplot}  Corner plot derived from our \texttt{dynesty} run towards the B1 region.}}
\end{center}
\end{figure}

\begin{figure}[h!]
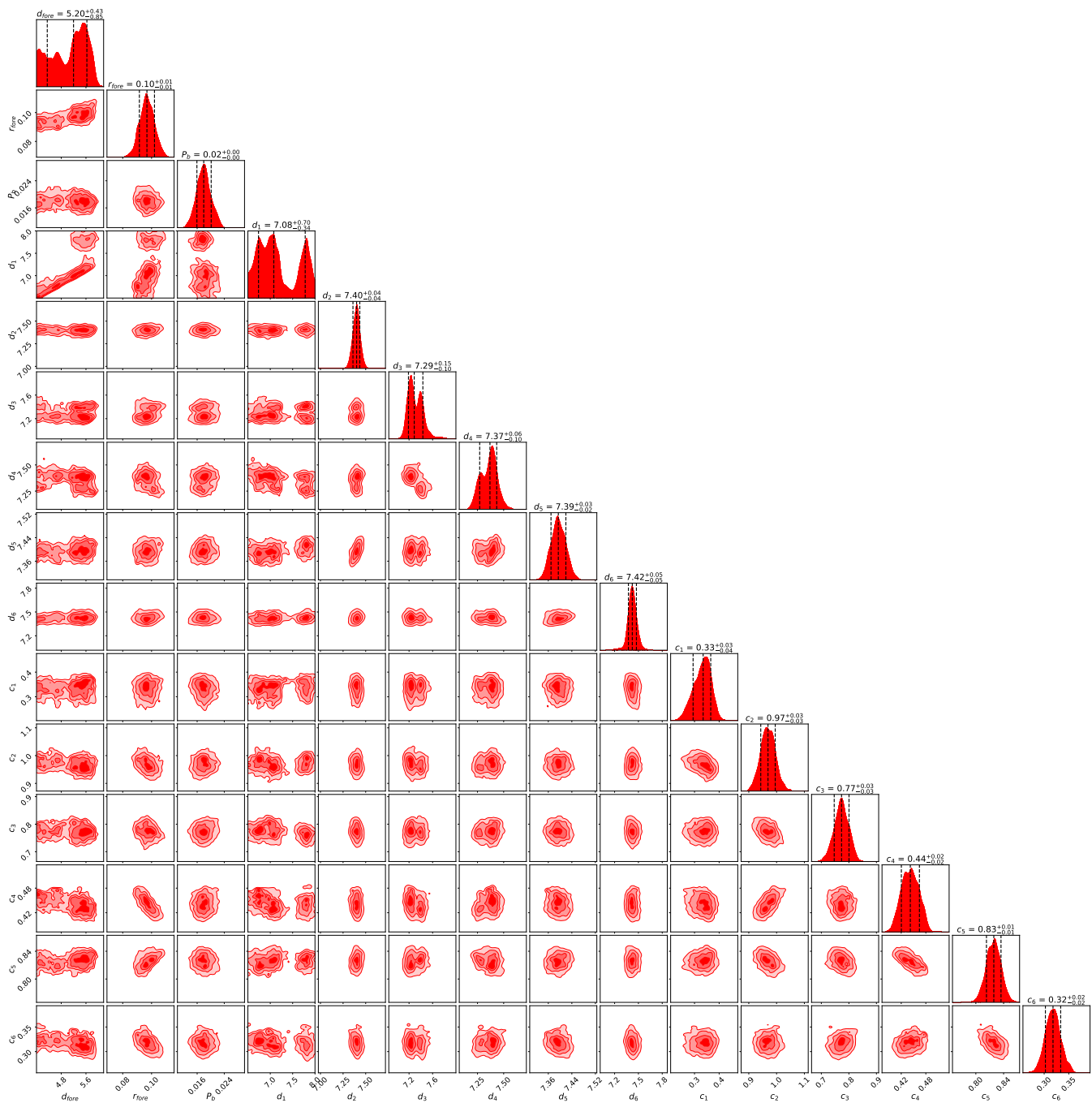

\begin{center}
\includegraphics[width=1.0\columnwidth]{{{NGC1333_outlier_FINALPUB_corner}}}
\caption{{\label{fig:NGC1333_cornerplot}  Corner plot derived from our \texttt{dynesty} run towards the NGC1333 region.}}
\end{center}
\end{figure}

\begin{figure}[h!]
\begin{center}
\includegraphics[width=1.0\columnwidth]{{{L1448_outlier_FINALPUB_corner}}}
\caption{{\label{fig:L1448_cornerplot}  Corner plot derived from our \texttt{dynesty} run towards the L1448 region.}}
\end{center}
\end{figure}

\begin{figure}[h!]
\begin{center}
\includegraphics[width=1.0\columnwidth]{{{L1451_outlier_FINALPUB_corner}}}
\caption{{\label{fig:L1451_cornerplot}  Corner plot derived from our \texttt{dynesty} run towards the L1451 region.}}
\end{center}
\end{figure}

\clearpage
\bibliographystyle{apj}
\bibliography{full_article_revision}

\begin{thebibliography}{}
\expandafter\ifx\csname natexlab\endcsname\relax\def\natexlab#1{#1}\fi

\bibitem[{{Arce} {et~al.}(2010){Arce}, {Borkin}, {Goodman}, {Pineda}, \&
  {Halle}}]{Arce_2010}
{Arce}, H.~G., {Borkin}, M.~A., {Goodman}, A.~A., {Pineda}, J.~E., \& {Halle},
  M.~W. 2010, \apj, 715, 1170

\bibitem[{{Bahcall} \& {Soneira}(1980)}]{Bahcall_1980}
{Bahcall}, J.~N., \& {Soneira}, R.~M. 1980, \apjs, 44, 73

\bibitem[{{Bally} {et~al.}(2008){Bally}, {Walawender}, {Johnstone}, {Kirk}, \&
  {Goodman}}]{Bally_2008}
{Bally}, J., {Walawender}, J., {Johnstone}, D., {Kirk}, H., \& {Goodman}, A.
  2008, {The Perseus Cloud}, ed. B.~{Reipurth}, 308

\bibitem[{{Beaumont} {et~al.}(2013){Beaumont}, {Offner}, {Shetty}, {Glover}, \&
  {Goodman}}]{Beaumont_2013}
{Beaumont}, C.~N., {Offner}, S. S.~R., {Shetty}, R., {Glover}, S. C.~O., \&
  {Goodman}, A.~A. 2013, \apj, 777, 173

\bibitem[{{Bohlin} {et~al.}(1978){Bohlin}, {Savage}, \& {Drake}}]{Bohlin_1978}
{Bohlin}, R.~C., {Savage}, B.~D., \& {Drake}, J.~F. 1978, \apj, 224, 132

\bibitem[{{Bressan} {et~al.}(2012){Bressan}, {Marigo}, {Girardi}, {Salasnich},
  {Dal Cero}, {Rubele}, \& {Nanni}}]{Bressan_2012}
{Bressan}, A., {Marigo}, P., {Girardi}, L., {et~al.} 2012, \mnras, 427, 127

\bibitem[{{Campbell} {et~al.}(2016){Campbell}, {Friesen}, {Martin}, {Caselli},
  {Kauffmann}, \& {Pineda}}]{Campbell_2016}
{Campbell}, J.~L., {Friesen}, R.~K., {Martin}, P.~G., {et~al.} 2016, \apj, 819,
  143

\bibitem[{\v{C}ernis(1990)}]{Cernis_1990}
{\v{C}}ernis, K. 1990, Ap\&SS, 166, 315

\bibitem[{{\v{C}}ernis(1993)}]{Cernis_1993}
---. 1993, BaltA, 2, 214

\bibitem[{{Chambers} {et~al.}(2016){Chambers}, {Magnier}, {Metcalfe},
  {Flewelling}, {Huber}, {Waters}, {Denneau}, {Draper}, {Farrow}, {Finkbeiner},
  {Holmberg}, {Koppenhoefer}, {Price}, {Saglia}, {Schlafly}, {Smartt},
  {Sweeney}, {Wainscoat}, {Burgett}, {Grav}, {Heasley}, {Hodapp}, {Jedicke},
  {Kaiser}, {Kudritzki}, {Luppino}, {Lupton}, {Monet}, {Morgan}, {Onaka},
  {Stubbs}, {Tonry}, {Banados}, {Bell}, {Bender}, {Bernard}, {Botticella},
  {Casertano}, {Chastel}, {Chen}, {Chen}, {Cole}, {Deacon}, {Frenk},
  {Fitzsimmons}, {Gezari}, {Goessl}, {Goggia}, {Goldman}, {Grebel}, {Hambly},
  {Hasinger}, {Heavens}, {Heckman}, {Henderson}, {Henning}, {Holman}, {Hopp},
  {Ip}, {Isani}, {Keyes}, {Koekemoer}, {Kotak}, {Long}, {Lucey}, {Liu},
  {Martin}, {McLean}, {Morganson}, {Murphy}, {Nieto-Santisteban}, {Norberg},
  {Peacock}, {Pier}, {Postman}, {Primak}, {Rae}, {Rest}, {Riess}, {Riffeser},
  {Rix}, {Roser}, {Schilbach}, {Schultz}, {Scolnic}, {Szalay}, {Seitz},
  {Shiao}, {Small}, {Smith}, {Soderblom}, {Taylor}, {Thakar}, {Thiel},
  {Thilker}, {Urata}, {Valenti}, {Walter}, {Watters}, {Werner}, {White},
  {Wood-Vasey}, \& {Wyse}}]{Chambers_2016}
{Chambers}, K.~C., {Magnier}, E.~A., {Metcalfe}, N., {et~al.} 2016, ArXiv
  e-prints, arXiv:1612.05560

\bibitem[{{Curtis} \& {Richer}(2010)}]{Curtis_2009}
{Curtis}, E.~I., \& {Richer}, J.~S. 2010, \mnras, 402, 603

\bibitem[{{Dame} {et~al.}(2001){Dame}, {Hartmann}, \& {Thaddeus}}]{Dame_2001}
{Dame}, T.~M., {Hartmann}, D., \& {Thaddeus}, P. 2001, \apj, 547, 792

\bibitem[{{de Zeeuw} {et~al.}(1999){de Zeeuw}, {Hoogerwerf}, {de Bruijne},
  {Brown}, \& {Blaauw}}]{de_Zeeuw_1999}
{de Zeeuw}, P.~T., {Hoogerwerf}, R., {de Bruijne}, J.~H.~J., {Brown}, A.~G.~A.,
  \& {Blaauw}, A. 1999, \aj, 117, 354

\bibitem[{{Enoch} {et~al.}(2006){Enoch}, {Young}, {Glenn}, {Evans}, {Golwala},
  {Sargent}, {Harvey}, {Aguirre}, {Goldin}, {Haig}, {Huard}, {Lange},
  {Laurent}, {Maloney}, {Mauskopf}, {Rossinot}, \& {Sayers}}]{Enoch_2006}
{Enoch}, M.~L., {Young}, K.~E., {Glenn}, J., {et~al.} 2006, \apj, 638, 293

\bibitem[{{Feroz} {et~al.}(2009){Feroz}, {Hobson}, \& {Bridges}}]{Feroz_2009}
{Feroz}, F., {Hobson}, M.~P., \& {Bridges}, M. 2009, \mnras, 398, 1601

\bibitem[{{Foreman-Mackey} {et~al.}(2013){Foreman-Mackey}, {Hogg}, {Lang}, \&
  {Goodman}}]{Foreman_Mackey_2013}
{Foreman-Mackey}, D., {Hogg}, D.~W., {Lang}, D., \& {Goodman}, J. 2013, \pasp,
  125, 306

\bibitem[{{Foster} {et~al.}(2013){Foster}, {Mandel}, {Pineda}, {Covey}, {Arce},
  \& {Goodman}}]{Foster_2013}
{Foster}, J.~B., {Mandel}, K.~S., {Pineda}, J.~E., {et~al.} 2013, \mnras, 428,
  1606

\bibitem[{{Gaia Collaboration} {et~al.}(2016){Gaia Collaboration}, {Prusti},
  {de Bruijne}, {Brown}, {Vallenari}, {Babusiaux}, {Bailer-Jones}, {Bastian},
  {Biermann}, {Evans}, \& et~al.}]{Gaia_2016}
{Gaia Collaboration}, {Prusti}, T., {de Bruijne}, J.~H.~J., {et~al.} 2016,
  \aap, 595, A1

\bibitem[{Goodman \& Weare(2010)}]{Goodman_2010}
Goodman, J., \& Weare, J. 2010, Communications in Applied Mathematics and
  Computational Science, 5, 65

\bibitem[{{Green} {et~al.}(2014){Green}, {Schlafly}, {Finkbeiner}, {Juri{\'c}},
  {Rix}, {Burgett}, {Chambers}, {Draper}, {Flewelling}, {Kudritzki}, {Magnier},
  {Martin}, {Metcalfe}, {Tonry}, {Wainscoat}, \& {Waters}}]{Green_2014}
{Green}, G.~M., {Schlafly}, E.~F., {Finkbeiner}, D.~P., {et~al.} 2014, \apj,
  783, 114

\bibitem[{{Green} {et~al.}(2015){Green}, {Schlafly}, {Finkbeiner}, {Rix},
  {Martin}, {Burgett}, {Draper}, {Flewelling}, {Hodapp}, {Kaiser}, {Kudritzki},
  {Magnier}, {Metcalfe}, {Price}, {Tonry}, \& {Wainscoat}}]{Green_2015}
---. 2015, \apj, 810, 25

\bibitem[{{Green} {et~al.}(2018){Green}, {Schlafly}, {Finkbeiner}, {Rix},
  {Martin}, {Burgett}, {Draper}, {Flewelling}, {Hodapp}, {Kaiser}, {Kudritzki},
  {Magnier}, {Metcalfe}, {Tonry}, {Wainscoat}, \& {Waters}}]{Green_2018}
{Green}, G.~M., {Schlafly}, E.~F., {Finkbeiner}, D., {et~al.} 2018, \mnras,
  478, 651

\bibitem[{{Handley} {et~al.}(2015){Handley}, {Hobson}, \&
  {Lasenby}}]{Handley_2015}
{Handley}, W.~J., {Hobson}, M.~P., \& {Lasenby}, A.~N. 2015, \mnras, 453, 4384

\bibitem[{{Harris} {et~al.}(1954){Harris}, {Morgan}, \& {Roman}}]{Harris_1954}
{Harris}, D.~L., {Morgan}, W.~W., \& {Roman}, N.~G. 1954, \apj, 119, 622

\bibitem[{{Higson} {et~al.}(2017){Higson}, {Handley}, {Hobson}, \&
  {Lasenby}}]{Higson_2017b}
{Higson}, E., {Handley}, W., {Hobson}, M., \& {Lasenby}, A. 2017, ArXiv
  e-prints, arXiv:1704.03459

\bibitem[{Higson {et~al.}(2018)Higson, Handley, Hobson, \&
  Lasenby}]{Higson_2017a}
Higson, E., Handley, W., Hobson, M., \& Lasenby, A. 2018, BayAn, 13, 873

\bibitem[{Hirota {et~al.}(2011)Hirota, Honma, Imai, Sunada, Ueno, Kobayashi, \&
  Kawaguchi}]{Hirota_2011}
Hirota, T., Honma, M., Imai, H., {et~al.} 2011, PASJ, 63, 1

\bibitem[{Hirota {et~al.}(2008)Hirota, Bushimata, Choi, Honma, Imai, Iwadate,
  Jike, Kameya, Kamohara, Kan-Ya, Kawaguchi, Kijima, Kobayashi, Kuji, Kurayama,
  Manabe, Miyaji, Nagayama, Nakagawa, Oh, Omodaka, Oyama, Sakai, Sasao, Sato,
  Shibata, Tamura, \& Yamashita}]{Hirota_2008}
Hirota, T., Bushimata, T., Choi, Y.~K., {et~al.} 2008, PASJ, 60, 37

\bibitem[{{Hogg} {et~al.}(2010){Hogg}, {Bovy}, \& {Lang}}]{Hogg_2010}
{Hogg}, D.~W., {Bovy}, J., \& {Lang}, D. 2010, ArXiv e-prints, arXiv:1008.4686

\bibitem[{{Huijser} {et~al.}(2015){Huijser}, {Goodman}, \&
  {Brewer}}]{Huijser_2015}
{Huijser}, D., {Goodman}, J., \& {Brewer}, B.~J. 2015, ArXiv e-prints,
  arXiv:1509.02230

\bibitem[{{Ivezi{\'c}} {et~al.}(2008){Ivezi{\'c}}, {Sesar}, {Juri{\'c}},
  {Bond}, {Dalcanton}, {Rockosi}, {Yanny}, {Newberg}, {Beers}, {Allende
  Prieto}, {Wilhelm}, {Lee}, {Sivarani}, {Norris}, {Bailer-Jones}, {Re
  Fiorentin}, {Schlegel}, {Uomoto}, {Lupton}, {Knapp}, {Gunn}, {Covey}, {Allyn
  Smith}, {Miknaitis}, {Doi}, {Tanaka}, {Fukugita}, {Kent}, {Finkbeiner},
  {Munn}, {Pier}, {Quinn}, {Hawley}, {Anderson}, {Kiuchi}, {Chen}, {Bushong},
  {Sohi}, {Haggard}, {Kimball}, {Barentine}, {Brewington}, {Harvanek},
  {Kleinman}, {Krzesinski}, {Long}, {Nitta}, {Snedden}, {Lee}, {Harris},
  {Brinkmann}, {Schneider}, \& {York}}]{Ivezic_2008}
{Ivezi{\'c}}, {\v Z}., {Sesar}, B., {Juri{\'c}}, M., {et~al.} 2008, \apj, 684,
  287

\bibitem[{{Juri{\'c}} {et~al.}(2008){Juri{\'c}}, {Ivezi{\'c}}, {Brooks},
  {Lupton}, {Schlegel}, {Finkbeiner}, {Padmanabhan}, {Bond}, {Sesar},
  {Rockosi}, {Knapp}, {Gunn}, {Sumi}, {Schneider}, {Barentine}, {Brewington},
  {Brinkmann}, {Fukugita}, {Harvanek}, {Kleinman}, {Krzesinski}, {Long},
  {Neilsen}, {Nitta}, {Snedden}, \& {York}}]{Juric_2008}
{Juri{\'c}}, M., {Ivezi{\'c}}, {\v Z}., {Brooks}, A., {et~al.} 2008, \apj, 673,
  864

\bibitem[{{Kirk} {et~al.}(2007){Kirk}, {Johnstone}, \& {Tafalla}}]{Kirk_2007}
{Kirk}, H., {Johnstone}, D., \& {Tafalla}, M. 2007, \apj, 668, 1042

\bibitem[{{Lee} {et~al.}(2015){Lee}, {Dunham}, {Myers}, {Tobin}, {Kristensen},
  {Pineda}, {Vorobyov}, {Offner}, {Arce}, {Li}, {Bourke}, {J{\o}rgensen},
  {Goodman}, {Sadavoy}, {Chandler}, {Harris}, {Kratter}, {Looney}, {Melis},
  {Perez}, \& {Segura-Cox}}]{Lee_2015}
{Lee}, K.~I., {Dunham}, M.~M., {Myers}, P.~C., {et~al.} 2015, \apj, 814, 114

\bibitem[{{Lindegren} {et~al.}(2018){Lindegren}, {Hern{\'a}ndez}, {Bombrun},
  {Klioner}, {Bastian}, {Ramos-Lerate}, {de Torres}, {Steidelm{\"u}ller},
  {Stephenson}, {Hobbs}, {Lammers}, {Biermann}, {Geyer}, {Hilger}, {Michalik},
  {Stampa}, {McMillan}, {Casta{\~n}eda}, {Clotet}, {Comoretto}, {Davidson},
  {Fabricius}, {Gracia}, {Hambly}, {Hutton}, {Mora}, {Portell}, {van Leeuwen},
  {Abbas}, {Abreu}, {Altmann}, {Andrei}, {Anglada}, {Balaguer-N{\'u}{\~n}ez},
  {Barache}, {Becciani}, {Bertone}, {Bianchi}, {Bouquillon}, {Bourda},
  {Br{\"u}semeister}, {Bucciarelli}, {Busonero}, {Buzzi}, {Cancelliere},
  {Carlucci}, {Charlot}, {Cheek}, {Crosta}, {Crowley}, {de Bruijne}, {de
  Felice}, {Drimmel}, {Esquej}, {Fienga}, {Fraile}, {Gai}, {Garralda},
  {Gonz{\'a}lez- Vidal}, {Guerra}, {Hauser}, {Hofmann}, {Holl}, {Jordan},
  {Lattanzi}, {Lenhardt}, {Liao}, {Licata}, {Lister}, {L{\"o}ffler},
  {Marchant}, {Martin-Fleitas}, {Messineo}, {Mignard}, {Morbidelli}, {Poggio},
  {Riva}, {Rowell}, {Salguero}, {Sarasso}, {Sciacca}, {Siddiqui}, {Smart},
  {Spagna}, {Steele}, {Taris}, {Torra}, {van Elteren}, {van Reeven}, \&
  {Vecchiato}}]{Lindegren_2018}
{Lindegren}, L., {Hern{\'a}ndez}, J., {Bombrun}, A., {et~al.} 2018, \aap, 616,
  A2

\bibitem[{{Lombardi}(2009)}]{Lombardi_2009}
{Lombardi}, M. 2009, \aap, 493, 735

\bibitem[{{Lombardi} \& {Alves}(2001)}]{Lombardi_2001}
{Lombardi}, M., \& {Alves}, J. 2001, \aap, 377, 1023

\bibitem[{{Lombardi} {et~al.}(2010){Lombardi}, {Lada}, \&
  {Alves}}]{Lombardi_2010}
{Lombardi}, M., {Lada}, C.~J., \& {Alves}, J. 2010, \aap, 512, A67

\bibitem[{{Magnier} {et~al.}(2016){Magnier}, {Schlafly}, {Finkbeiner}, {Tonry},
  {Goldman}, {R{\"o}ser}, {Schilbach}, {Chambers}, {Flewelling}, {Huber},
  {Price}, {Sweeney}, {Waters}, {Denneau}, {Draper}, {Hodapp}, {Jedicke},
  {Kudritzki}, {Metcalfe}, {Stubbs}, \& {Wainscoast}}]{Magnier_2016}
{Magnier}, E.~A., {Schlafly}, E.~F., {Finkbeiner}, D.~P., {et~al.} 2016, ArXiv
  e-prints, arXiv:1612.05242

\bibitem[{{Maureira} {et~al.}(2017){Maureira}, {Arce}, {Dunham}, {Pineda},
  {Fern{\'a}ndez-L{\'o}pez}, {Chen}, \& {Mardones}}]{Maureira_2017}
{Maureira}, M.~J., {Arce}, H.~G., {Dunham}, M.~M., {et~al.} 2017, \apj, 838, 60

\bibitem[{{Ortiz-Le{\'o}n} {et~al.}(2018){Ortiz-Le{\'o}n}, {Loinard}, {Dzib},
  {Galli}, {Kounkel}, {Mioduszewski}, {Rodr{\'\i}guez}, {Torres}, {Hartmann},
  {Boden}, {Evans}, {Brice{\~n}o}, \& {Tobin}}]{Ortiz_Leon_2018}
{Ortiz-Le{\'o}n}, G.~N., {Loinard}, L., {Dzib}, S.~A., {et~al.} 2018, \apj,
  865, 73

\bibitem[{{Pineda} {et~al.}(2008){Pineda}, {Caselli}, \&
  {Goodman}}]{Pineda_2008}
{Pineda}, J.~E., {Caselli}, P., \& {Goodman}, A.~A. 2008, \apj, 679, 481

\bibitem[{{Pineda} {et~al.}(2011){Pineda}, {Arce}, {Schnee}, {Goodman},
  {Bourke}, {Foster}, {Robitaille}, {Tanner}, {Kauffmann}, {Tafalla},
  {Caselli}, \& {Anglada}}]{Pineda_2011}
{Pineda}, J.~E., {Arce}, H.~G., {Schnee}, S., {et~al.} 2011, \apj, 743, 201

\bibitem[{{Planck Collaboration} {et~al.}(2011){Planck Collaboration}, {Ade},
  {Aghanim}, {Arnaud}, {Ashdown}, {Aumont}, {Baccigalupi}, {Balbi}, {Banday},
  {Barreiro}, \& et~al.}]{Planck_2011}
{Planck Collaboration}, {Ade}, P.~A.~R., {Aghanim}, N., {et~al.} 2011, \aap,
  536, A19

\bibitem[{{Rebull} {et~al.}(2007){Rebull}, {Stapelfeldt}, {Evans},
  {J{\o}rgensen}, {Harvey}, {Brooke}, {Bourke}, {Padgett}, {Chapman}, {Lai},
  {Spiesman}, {Noriega-Crespo}, {Mer{\'{\i}}n}, {Huard}, {Allen}, {Blake},
  {Jarrett}, {Koerner}, {Mundy}, {Myers}, {Sargent}, {van Dishoeck}, {Wahhaj},
  \& {Young}}]{Rebull_2007}
{Rebull}, L.~M., {Stapelfeldt}, K.~R., {Evans}, II, N.~J., {et~al.} 2007,
  \apjs, 171, 447

\bibitem[{Ridge {et~al.}(2006)Ridge, Francesco, Kirk, Li, Goodman, Alves, Arce,
  Borkin, Caselli, Foster, Heyer, Johnstone, Kosslyn, Lombardi, Pineda, Schnee,
  \& Tafalla}]{Ridge_2006}
Ridge, N.~A., Francesco, J.~D., Kirk, H., {et~al.} 2006, AJ, 131, 2921

\bibitem[{{Robin} {et~al.}(2003){Robin}, {Reyl{\'e}}, {Derri{\`e}re}, \&
  {Picaud}}]{Robin_2003}
{Robin}, A.~C., {Reyl{\'e}}, C., {Derri{\`e}re}, S., \& {Picaud}, S. 2003,
  \aap, 409, 523

\bibitem[{{Roman-Duval} {et~al.}(2009){Roman-Duval}, {Jackson}, {Heyer},
  {Johnson}, {Rathborne}, {Shah}, \& {Simon}}]{Roman_Duval_2009}
{Roman-Duval}, J., {Jackson}, J.~M., {Heyer}, M., {et~al.} 2009, \apj, 699,
  1153

\bibitem[{{Rosolowsky} {et~al.}(2008){Rosolowsky}, {Pineda}, {Foster},
  {Borkin}, {Kauffmann}, {Caselli}, {Myers}, \& {Goodman}}]{Rosolowsky_2008}
{Rosolowsky}, E.~W., {Pineda}, J.~E., {Foster}, J.~B., {et~al.} 2008, \apjs,
  175, 509

\bibitem[{{Sadavoy} {et~al.}(2014){Sadavoy}, {Di Francesco}, {Andr{\'e}},
  {Pezzuto}, {Bernard}, {Maury}, {Men'shchikov}, {Motte}, {Nguyen-Lu'o'ng},
  {Schneider}, {Arzoumanian}, {Benedettini}, {Bontemps}, {Elia}, {Hennemann},
  {Hill}, {K{\"o}nyves}, {Louvet}, {Peretto}, {Roy}, \& {White}}]{Sadavoy_2014}
{Sadavoy}, S.~I., {Di Francesco}, J., {Andr{\'e}}, P., {et~al.} 2014, \apjl,
  787, L18

\bibitem[{Schlafly {et~al.}(2014)Schlafly, Green, Finkbeiner, Rix, Bell,
  Burgett, Chambers, Draper, Hodapp, Kaiser, Magnier, Martin, Metcalfe, Price,
  \& Tonry}]{Schlafly_2014}
Schlafly, E.~F., Green, G., Finkbeiner, D.~P., {et~al.} 2014, ApJ, 786, 29

\bibitem[{{Schlafly} {et~al.}(2016){Schlafly}, {Meisner}, {Stutz},
  {Kainulainen}, {Peek}, {Tchernyshyov}, {Rix}, {Finkbeiner}, {Covey}, {Green},
  {Bell}, {Burgett}, {Chambers}, {Draper}, {Flewelling}, {Hodapp}, {Kaiser},
  {Magnier}, {Martin}, {Metcalfe}, {Wainscoat}, \& {Waters}}]{Schlafly_2016}
{Schlafly}, E.~F., {Meisner}, A.~M., {Stutz}, A.~M., {et~al.} 2016, \apj, 821,
  78

\bibitem[{{Shetty} {et~al.}(2011){Shetty}, {Glover}, {Dullemond}, \&
  {Klessen}}]{Shetty_2011}
{Shetty}, R., {Glover}, S.~C., {Dullemond}, C.~P., \& {Klessen}, R.~S. 2011,
  \mnras, 412, 1686

\bibitem[{Skilling(2006)}]{Skilling_2006}
Skilling, J. 2006, BayAn, 1, 833

\bibitem[{{Skrutskie} {et~al.}(2006){Skrutskie}, {Cutri}, {Stiening},
  {Weinberg}, {Schneider}, {Carpenter}, {Beichman}, {Capps}, {Chester},
  {Elias}, {Huchra}, {Liebert}, {Lonsdale}, {Monet}, {Price}, {Seitzer},
  {Jarrett}, {Kirkpatrick}, {Gizis}, {Howard}, {Evans}, {Fowler}, {Fullmer},
  {Hurt}, {Light}, {Kopan}, {Marsh}, {McCallon}, {Tam}, {Van Dyk}, \&
  {Wheelock}}]{Skrutskie_2006}
{Skrutskie}, M.~F., {Cutri}, R.~M., {Stiening}, R., {et~al.} 2006, \aj, 131,
  1163

\bibitem[{{Storm} {et~al.}(2016){Storm}, {Mundy}, {Lee},
  {Fern{\'a}ndez-L{\'o}pez}, {Looney}, {Teuben}, {Arce}, {Rosolowsky},
  {Meisner}, {Isella}, {Kauffmann}, {Shirley}, {Kwon}, {Plunkett}, {Pound},
  {Segura-Cox}, {Tassis}, {Tobin}, {Volgenau}, {Crutcher}, \&
  {Testi}}]{Storm_2016}
{Storm}, S., {Mundy}, L.~G., {Lee}, K.~I., {et~al.} 2016, \apj, 830, 127

\bibitem[{{Tchernyshyov} \& {Peek}(2017)}]{Tchernyshyov_2017}
{Tchernyshyov}, K., \& {Peek}, J.~E.~G. 2017, \aj, 153, 8

\bibitem[{{Tchernyshyov} {et~al.}(2018){Tchernyshyov}, {Peek}, \&
  {Zasowski}}]{Tchernyshyov_2018}
{Tchernyshyov}, K., {Peek}, J.~E.~G., \& {Zasowski}, G. 2018, ArXiv e-prints,
  arXiv:1808.01286

\bibitem[{{Tritsis} \& {Tassis}(2018)}]{Tassis_2018}
{Tritsis}, A., \& {Tassis}, K. 2018, Sci, 360, 635

\bibitem[{{Trullols} \& {Jordi}(1997)}]{Trullols_1997}
{Trullols}, E., \& {Jordi}, C. 1997, \aap, 324, 549

\bibitem[{{{\v C}ernis} \& {Strai{\v z}ys}(2003)}]{Cernis_2003}
{{\v C}ernis}, K., \& {Strai{\v z}ys}, V. 2003, BaltA, 12, 301

\bibitem[{{Warren} {et~al.}(2007){Warren}, {Cross}, {Dye}, {Hambly}, {Almaini},
  {Edge}, {Hewett}, {Hodgkin}, {Irwin}, {Jameson}, {Lawrence}, {Lucas},
  {Mortlock}, {Adamson}, {Bryant}, {Collins}, {Davis}, {Emerson}, {Evans},
  {Gonzales-Solares}, {Hirst}, {Kerr}, {Lewis}, {Mann}, {Rawlings}, {Read},
  {Riello}, {Sutorius}, \& {Varricatt}}]{Warren_2007}
{Warren}, S.~J., {Cross}, N.~J.~G., {Dye}, S., {et~al.} 2007, ArXiv
  Astrophysics e-prints, astro-ph/0703037

\end{thebibliography}

\end{document}